\documentclass[
reprint,
amsmath,amssymb,
aps,
]{revtex4-1}

\usepackage{amsmath}
\usepackage{slashed}
\usepackage{contour}
\usepackage{graphicx}
\usepackage{mathtools}
\usepackage{tikz}
\usepackage{amsmath}
\usetikzlibrary{arrows,shapes}
\usepackage{xcolor}
\usepackage{amsmath,amssymb,latexsym}
\usepackage{enumerate}
\usepackage{comment}

\newcommand{\red}[1]{{\color{red} #1}}
\newcommand{\blue}[1]{{\color{blue} #1}}

\begin{document}

\title{Research Notes:\\Gradient sensing in Bayesian chemotaxis}

\author{Andrea Auconi$^{1}$}
\author{Maja Novak$^{1,2}$}
\author{Benjamin M. Friedrich$^{1,3}$}
\affiliation{$1$ cfaed, Technische Universit\"at Dresden, 01069 Dresden, Germany}
\affiliation{$2$ Department of Physics, Faculty of Science, University of Zagreb, Bijenička cesta 32, 10000 Zagreb, Croatia}
\affiliation{$3$ Cluster of Excellence “Physics of Life” - 01307 Dresden, Germany}

\date{\today}

\begin{abstract}
	Bayesian chemotaxis is an information-based target search problem inspired by biological chemotaxis. It is defined by a decision strategy coupled to the dynamic estimation of target position from detections of signaling molecules. We extend the case of a point-like agent previously introduced in [Vergassola et al., Nature 2007], which establishes concentration sensing as the dominant contribution to information processing, to the case of a circular agent of small finite size. We identify gradient sensing and a Laplacian correction to concentration sensing as the two leading-order expansion terms in the expected entropy variation. Numerically, we find that the impact of gradient sensing is most relevant because it provides direct directional information to break symmetry in likelihood distributions, which are generally circle-shaped by concentration sensing. 
\end{abstract}

\maketitle

\section{Part I: Statement of the problem and summary of results}

\paragraph{Introduction.}

Biological cells and organisms navigate their environments, e.g., in search for nutrient sources, guided by the detection of signaling molecules \cite{bialek2012biophysics, alon1999robustness, friedrich2007chemotaxis}. Knowledge of the time and location of such detections, typically receptor-ligand binding events, provides information about the target location if the target is a source of ligands. More precisely, an agent can reduce its uncertainty about target location if the binding events time series is interpreted with respect to a \textit{model} of the environment. In other words, a chemotactic agent should know something about the relation between binding statistics and target location in order to process the information from binding events.

The estimation of a time-varying hidden state from noisy measurements is the subject of stochastic filtering theory \cite{fujisaki1972stochastic, rogers2000diffusions, petris2009dynamic}. The celebrated Kalman filter provides the analytical form of the estimation update scheme for Gaussian systems, and it was recently applied to biological examples such as concentration and direction sensing in the linear regime \cite{husain2019kalman, mora2019physical, novak2021bayesian, malaguti2021theory}.
Chemotaxis can be formalized as stochastic filtering coupled to a decision making problem, namely the decision of where to move next.

Just knowing that the binding rate increases towards the target is sufficient for a simple strategy of gradient ascent, where the agent only needs to repeatedly estimate the local gradient from the binding asymmetry across its diameter \cite{devreotes1988chemotaxis}, or across short-distance runs like in bacterial run-and-tumble chemotaxis \cite{segall1986temporal}, referred to as spatial or temporal comparison in the biological literature, respectively.

However, if concentrations are dilute and therefore binding events are sparse, then the estimation of gradients becomes very slow and inefficient. Nevertheless, if an agent is equipped with a model of the binding rate as a function of the relative target location, then it can gain information also by estimating the local binding rate. Indeed, the signal-to-noise ratio in the estimation of a concentration from binding events is typically much larger than that of a gradient.

The chemo-sensory navigation of a point-like agent performing stochastic filtering on a likelihood map was discussed in the pioneering work of Vergassola et al. \cite{vergassola2007infotaxis, masson2009chasing}, where the authors also introduced a decision strategy that consists in maximizing the expected information gain locally in time, there named \textit{infotaxis}. Analytical expressions for the information gain and decision making were subsequently given in \cite{barbieri2011trajectories}. We would like to refer to this and similar search problems as Bayesian chemotaxis.

Intuitively, a point-like agent can estimate the local gradient, but only indirectly by movement and temporal comparison of estimated concentrations. On the contrary, an agent with small finite size can also obtain direct information on the local gradient, as the location of a single binding event on its surface is itself a directional cue \cite{hu2010physical, novak2021bayesian}.
Therefore, we interpret the case of point-like agent as pure concentration sensing, while the small finite-size agent performs simultaneously concentration and gradient sensing.

In these Research Notes, we introduce gradient sensing from an information gain expansion with respect to agent size. The zeroth order terms are the motility noise, and the concentration sensing discussed in \cite{vergassola2007infotaxis,masson2009chasing,barbieri2011trajectories}. The second order of the expansion gives the gradient sensing term, as well as a Laplacian correction to concentration sensing which, to the best of our knowledge, was not yet discussed in the literature. We evaluate the relative importance of these terms in a steady-state chemotactic search model in two space dimensions. We discuss the increase on target search efficiency deriving from a finite agent size. Finally we briefly discuss the decision process comparing infotaxis to the simpler maximum likelihood and minimum distance direction strategies.

\paragraph{The model.}
A disk-shaped agent of radius $a$ moving in 2D space is immersed in a concentration profile of signaling molecules released by a target. These molecules are detected on the agent's circumference through binding events happening at a rate proportional to the local concentration. Let us use a coordinate system centered at the agent, and let $r(\mathbf{x})$ denote the binding rate at the agent position $\mathbf{0}$ when the target is at position $\mathbf{x}$. 
We denote by $p(\mathbf{x}_t)$ the smooth likelihood distribution formalizing the agent's knowledge about target location at time $t$, which is based on previous measurements and dynamics up to time $t$. The agent moves with a velocity $\mathbf{v}_t$ having a constant speed $v$, and a direction angle $\theta_t$, which is continuously decided based on the current likelihood. The target further undergoes translational diffusion with constant diffusion coefficient $D$.

\paragraph{Stochastic filtering.}
The likelihood $p(\mathbf{x}_t)$ is continuously updated in response to the binding events time series and to account for the target stochastic dynamics in a Bayesian inference scheme called stochastic filtering. Consider a time interval $[0,\tau]$ with $\tau>0$ sufficiently small so that variations of the binding rate due to movement within $[0,\tau]$ are negligible. Let us further assume that the agent velocity in real space follows a protocol decided at $t=0$. The evolution of the likelihood is then decomposed into a prediction step $p(\mathbf{x}_{\tau})=\int d\mathbf{x}_0\, p(\mathbf{x}_0) p(\mathbf{x}_{\tau}|\mathbf{x}_0)$, which is driven by the transition density of the target position $p(\mathbf{x}_{\tau}|\mathbf{x}_0)$, and of an update step
$p(\mathbf{x}_{\tau}|\mathbf{m}_{\tau})=p(\mathbf{x}_{\tau}) p(\mathbf{m}_{\tau}|\mathbf{x}_{\tau}) / p(\mathbf{m}_{\tau})$,
which is driven by the measurement $\mathbf{m}_{\tau}$, that here is a record of the number and locations on the agent's circumference of the binding events happened within the interval $[0,\tau)$. The measurement model $p(\mathbf{m}_{\tau}|\mathbf{x}_{\tau})$ formalizes the relation between binding rate and target location, which is known by the agent in Bayesian chemotaxis. The current measurement probability $p(\mathbf{m}_{\tau})=\int d\mathbf{x}_{\tau}\, p(\mathbf{x}_{\tau}) p(\mathbf{m}_{\tau}|\mathbf{x}_{\tau})$ formalizes the agent's uncertainty at time $0$ about the outcome of the measurement $\mathbf{m}_{\tau}$, which is delivered to it at time $\tau$. In the limit $\tau\rightarrow 0$, the Bayesian updating is integrated in the non-anticipating Ito stochastic calculus scheme to obtain a time-continuous inference and decision process, see Section II of these Notes.

\paragraph{Expansion in the agent's size.}
If we neglect the effect of the agent body and movement on the sensing field, the inference problem is equivalent to that made by a density of point-like agents distributed on the circumference $a\{\mathbf{e^{i\theta}}\}_{\theta=[0,2\pi)}$, where $\mathbf{e^{i\theta}}$ is a unit vector in 2D with direction defined by the angle $\theta$ from a fixed axis in the agent's reference frame.
Accordingly, the event probability density per unit angle in a small time interval $\tau$ is $p(\mathbf{m}_{\tau}= \mathbf{e^{i\theta}}|\mathbf{x})=\frac{\tau}{2\pi} r(\mathbf{x}-a\mathbf{e^{i\theta}})+\mathcal{O}(\tau^2)$, where the normalization is chosen so that in the limit $a\rightarrow 0$ one recovers the case of a single point-like agent.
Note that the measurement $\mathbf{m}_{\tau}$ is  vector-valued, assuming $\mathbf{0}$ (no events) almost surely for $\tau\rightarrow 0$, and $\mathbf{e^{i\theta}}$ in the case of a binding event at angle $\theta$.
If the binding rate differences along the agent's body are small compared to the absolute binding rate, then we can study the effect of the agent size on stochastic filtering through a second order Taylor expansion of the binding rate. For an event at angle $\theta$, we write
\begin{multline}\label{binding rate expansion}
r(\mathbf{x}-a\mathbf{e^{i\theta}}) = r(\mathbf{x}) -a\,\mathbf{e^{i\theta}} \cdot \mathbf{\nabla} r(\mathbf{x}) \\+\frac{1}{2} a^2 \mathbf{e^{i\theta}}\cdot\left( \mathbf{H}(\mathbf{x}) \,\mathbf{e^{i\theta}} \right) +\mathcal{O}(a^3) ,
\end{multline}
where $ \mathbf{H}(\mathbf{x})\equiv \mathbf{\nabla} \otimes \mathbf{\nabla} r(\mathbf{x})$ is the Hessian matrix of the binding rate evaluated at $\mathbf{x}$. Using the expansion Eq. \eqref{binding rate expansion} in the stochastic filtering problem, we obtain the corresponding expansion of the likelihood stochastic evolution equation, whose expression is given in Section II.

\paragraph{Entropy dynamics}
We quantify the agent's current uncertainty about target location by the Shannon entropy $S[p(\mathbf{x})]= - \int d\mathbf{x}\, p(\mathbf{x}) \ln p(\mathbf{x})$.
The expected entropy variation up to second order in $a$ reads
\begin{multline}\label{E dS}
\frac{\left\langle dS \right\rangle}{dt} 
= -D\left\langle  \mathbf{\nabla}^2 \ln p \right\rangle  
+  \left\langle \left(r +\frac{a^2}{4}\nabla^2 r \right) \ln \left(\frac{\left\langle  r\right\rangle}{r}\right) \right\rangle \\ 
+\frac{a^2}{4}  \left( \frac{|| \langle\mathbf{\nabla} r \rangle ||^2}{\langle r \rangle}-\left\langle \frac{|| \mathbf{\nabla} r ||^2}{r}  \right\rangle  \right) +\mathcal{O}(a^3),
\end{multline}
where $||\cdot||$ denotes Euclidean norm, the averaging brackets are meant with respect to the current likelihood $p$, and we dropped the explicit dependence on $\mathbf{x}$. Equation \eqref{E dS}, whose derivation is given in Section II, is the main result of these Research Notes. Let us now discuss its physical meaning.

The first zeroth order term, $-D\left\langle  \mathbf{\nabla}^2 \ln p \right\rangle\geq 0$, is the information erasure due to translational diffusion, and it is always nonnegative having the form of a Fisher information \cite{amari2016information}. The second zeroth order term, $\left\langle r \ln  (  \langle  r\rangle /r  ) \right\rangle \leq 0$, is the concentration sensing already discussed in \cite{vergassola2007infotaxis,barbieri2011trajectories}, and it is always nonpositive by Jensen's inequality.

We identify the second order term $ || \langle\mathbf{\nabla} r \rangle ||^2/\langle r \rangle -\left\langle || \mathbf{\nabla} r ||^2/r  \right\rangle \leq 0 $ as \textit{gradient sensing}, and it is always nonpositive having the form of the expectation of a negative square. Indeed, once a binding event has happened, the expected information gain from further knowing its location on the circumference can only be positive, being a type of conditional mutual information.
Note that for this term, if we consider an agent who periodically erases all of its current directional knowledge so that $\langle\mathbf{\nabla} r \rangle =\mathbf{0}$, then we obtain simply $\left\langle || \mathbf{\nabla} r ||^2/r  \right\rangle$. Depending on whether additionally distance information becomes erased, this expectation value is either taken with respect to a fixed prior or and adaptive one \cite{ito2015maxwell}.

The other second order term $\left\langle \left(\nabla^2 r \right) \ln \left(\left\langle  r\right\rangle/r\right) \right\rangle$ is interpreted as a \textit{Laplacian correction} to concentration sensing. This is due to the convexity of the binding rate field, which is the leading order correction in the total binding rate on the agent's circumference, $\int_0^{2\pi}(d\theta/2\pi)\, r(\mathbf{x}+ a\mathbf{e^{i\theta}}) = r(\mathbf{x})+(a^2/4) \nabla^2 r(\mathbf{x})+\mathcal{O}(a^4) $. The impact of the Laplacian correction on the entropy variation can be positive or negative depending on the particular likelihood configuration and binding rate field shape.  

From Eq. \eqref{E dS}, we can derive as limiting cases two known examples of stochastic filtering in biology, which are mathematically analogous to the Kalman filter \cite{husain2019kalman, petris2009dynamic}.

\paragraph{Limiting cases.}

We start with pure direction sensing of \cite{novak2021bayesian}. Consider a target constrained at a fixed position at distance $R$, which generates a conic field $r(||\mathbf{x}||)=r_0-||\mathbf{\nabla}r||\,||\mathbf{x}||$, with $r_0>0$, $||\mathbf{\nabla}r||>0$. The agent does not move but undergoes rotational diffusion with coefficient $D_\mathrm{rot}$, defined as Brownian motion in the angular variable $\theta$. In the small noise limit $D_\mathrm{rot}\rightarrow 0$ the angular likelihood distribution is Gaussian with mean $\mu_{\theta}$ and variance $\sigma^2_{\theta}$.  Concentration sensing is irrelevant here because we can just substitute $r = r(R)>0$, so that only gradient sensing and rotational diffusion contribute to the entropy variation. The entropy variation due to rotational diffusion is a type of Fisher information \cite{amari2016information}, and in the linear regime the corresponding Cramer-Rao bound $- \left\langle  \partial^2_{\theta} \ln p \right\rangle\geq 1/\sigma^2_{\theta}$ is saturated, as can be immediately seen by taking the $x$ axis along the direction of the current estimate $\mu_{\theta}$, and linearizing $y=R\theta +\mathcal{O}(\theta^2)$. Similarly, the gradient sensing is found by linearizing $|| \langle\mathbf{\nabla} r \rangle ||^2 =  ||\mathbf{\nabla}r||^2 (1-\sigma^2_{\theta}) +\mathcal{O}(\sigma^3_{\theta})$. Then the angular variance which makes the expected entropy variation vanish satisfies
\begin{equation}
 \sigma^2_{\theta}\big|_{\langle dS\rangle=0} = \sqrt{D_\mathrm{rot}}\, \frac{2\sqrt{r(R)}}{a ||\mathbf{\nabla}r||}.
\end{equation}
We see that the total rate of events $r(R)$ acts as Poissonian noise for the signal $2a||\mathbf{\nabla}r||$, which is the rate gradient along the agent body. The angular variance scales as the square-root of the rotational noise $D_\mathrm{rot}$.

Similarly, we derive the pure concentration sensing case of \cite{mora2019physical} by considering Eq. \eqref{E dS} in 1D, for a point-like agent ($a=0$) with translational diffusion whose likelihood at $t=0$ is non-zero only on one side, say for $x>0$. We take again the linear binding rate $r(x)=r_0- |\mathbf{\nabla}r| |x|$ for simplicity, and expand the concentration sensing to obtain
$
 \sigma^2_{x}\big|_{\langle dS\rangle=0} = \sqrt{D}\, \frac{2\sqrt{r(R)}}{|\mathbf{\nabla}r|},
$
which also scales as the square-root of motility noise.

\paragraph{Gradient sensing in steady-state Bayesian chemotaxis.}

To evaluate the relative importance of the expansion terms of Eq. \eqref{E dS} in the full Bayesian chemotaxis model, we plot the histogram of their realized values over the steady-state dynamics obtained with a binding rate field $r(\mathbf{x})=\lambda ||\mathbf{x}||^{-1}$, with $\lambda>0$, whose particular form is taken for numerical convenience. Here, we already employ a decision strategy (infotaxis) detailed in the next section. To ensure the existence of an ergodic steady-state dynamics for the chemotaxis process, we impose walls for the relative target distance $R_\mathrm{min}<||\mathbf{x}||<R_\mathrm{max}$. The lower bound $R_\mathrm{min}$, which can be interpreted as the target size, is taken for numerical stability and to ensure that the second order expansion holds in the allowed region.

\begin{center}
\begin{figure}
     \includegraphics[scale=0.56]{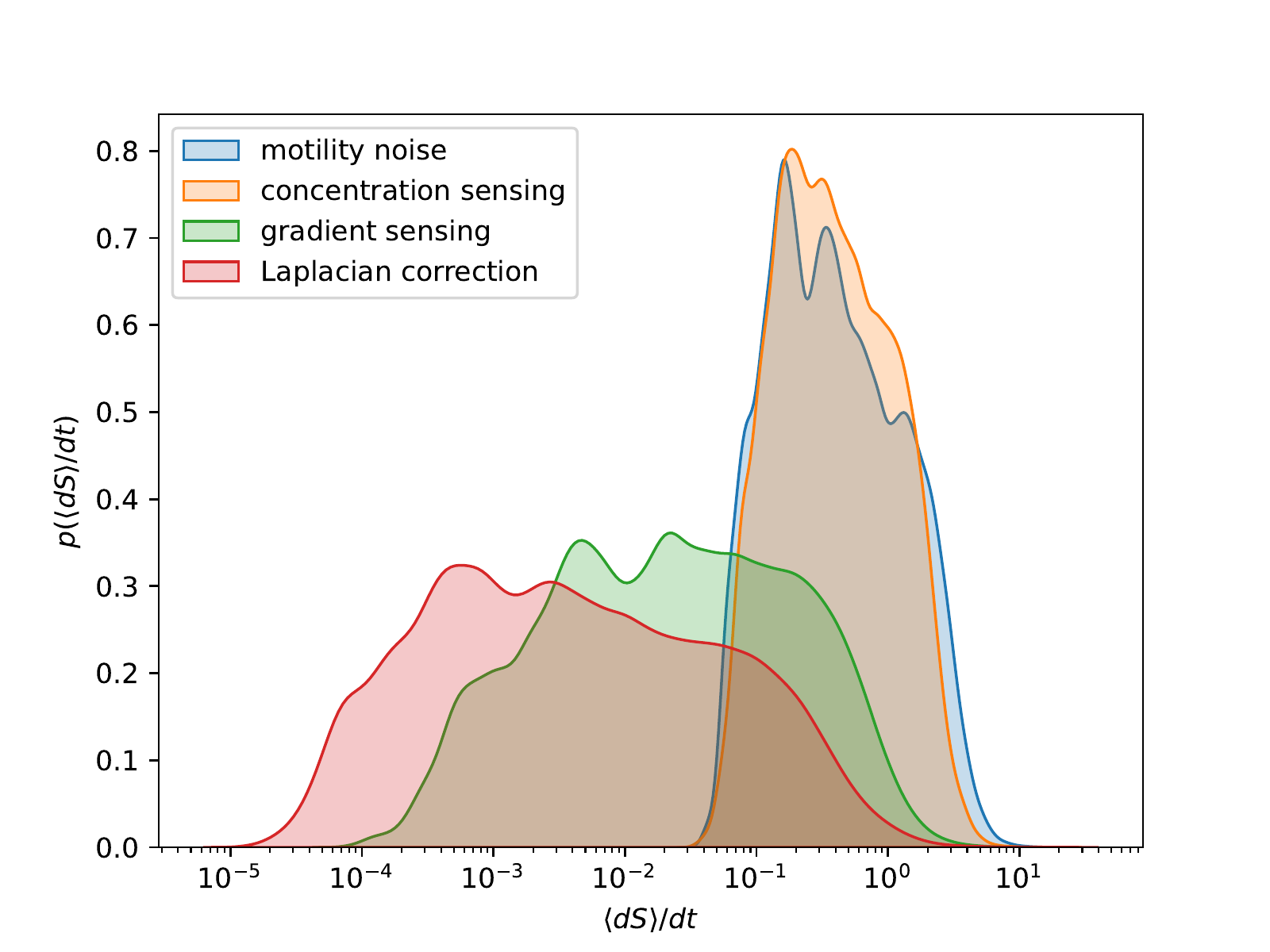}
	\caption{Distribution of the absolute values of the four terms in the expected entropy variation equation, Eq. \eqref{E dS}, evaluated in a steady-state dynamics with parameters $a=0.01$, $\lambda=2$, $v=0.01$, $D=2.5 \cdot 10^{-4}$, $R_\mathrm{min}=0.03$, $R_\mathrm{max}=0.87$. Note that the motility noise is positive, while the measurement components are all negative.}
	\label{Fig1}
\end{figure}
\end{center}

\begin{center}
\begin{figure}
     \includegraphics[scale=0.58, trim = {0.6cm 50 1 50}, clip]{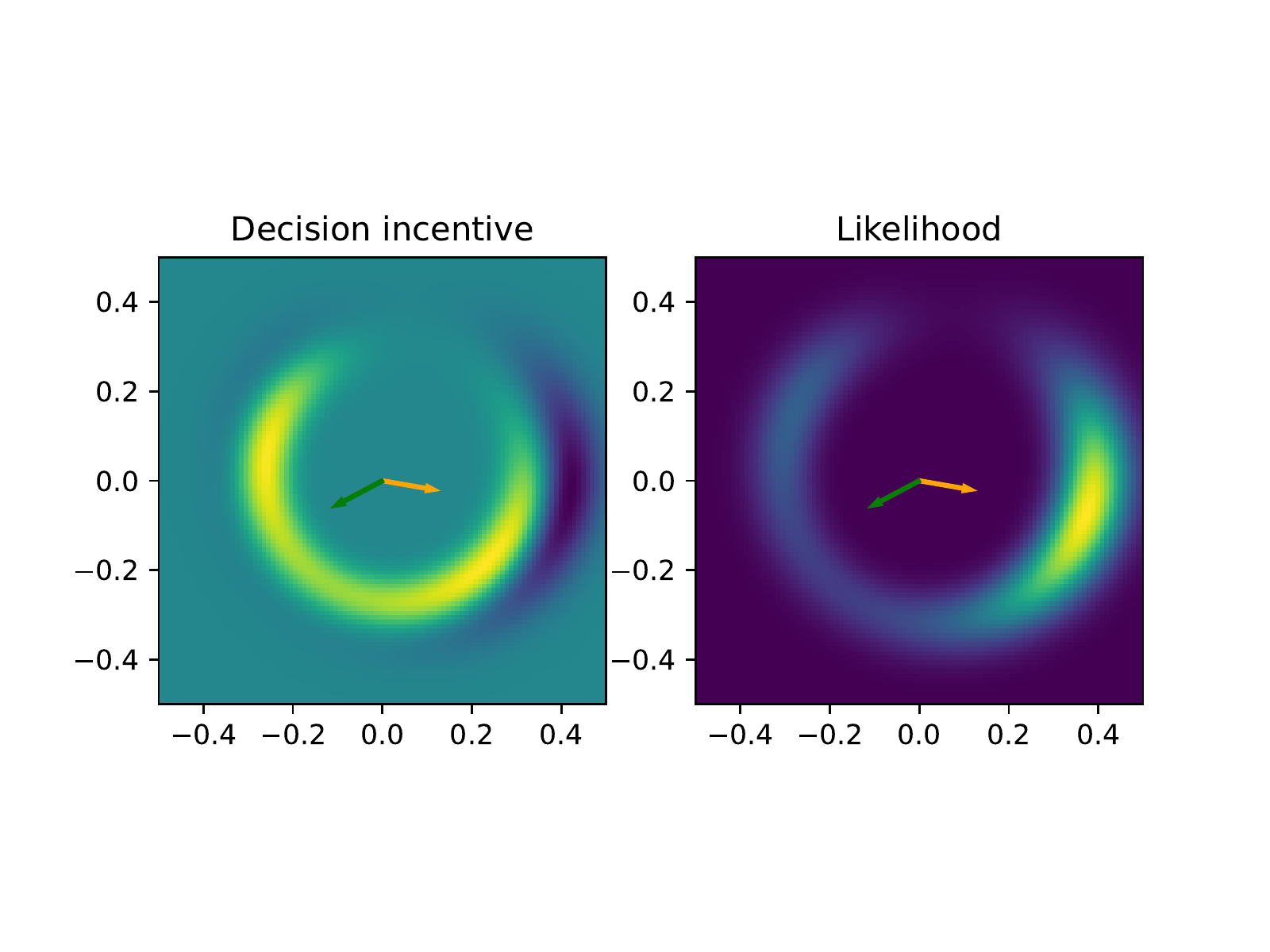}
\caption{Typical likelihood configuration $p$, and corresponding decision incentive $q$, see Eq. \eqref{infotaxis a} and following discussion. Dark blue regions in the decision incentive plot correspond to $q<0$, i.e., are repulsive. The orange arrow corresponds to the maximum likelihood direction, while the green arrow to the infotaxis direction. The particular configuration was selected from a long steady-state realization with parameters $a=0.01$, $\lambda=2$, $v=0.01$, $D=2.5 \cdot 10^{-4}$, $R_\mathrm{min}=0.03$, $R_\mathrm{max}=0.87$.}
	\label{Fig2}
\end{figure}
\end{center}

We find that gradient sensing is of larger magnitude compared to the Laplacian correction, see Fig. (\ref{Fig1}). This is understood looking at the typical likelihood shapes, see Fig. (\ref{Fig2}), (\ref{shapes}). The main drivers of the likelihood dynamics are the velocity-induced translation, the diffusion-induced smoothing, and the concentration sensing. In particular, concentration sensing gives rise to circle-shaped distributions, because we consider a concentration field which is symmetric around the target and the agent knows the relation between binding rate and distance. The likelihood is never an exact circle but rather an annulus whose width is increased by motility noise (effective diffusion of the target). Such annuli are then transformed into semicircles, bimodal, and single peak distributions as a result of the dynamics and measurements with efficient decision strategies.

In particular, bimodal distributions can arise from an annulus when the agent moves to a new position, 
where a new measurement effectively amounts to an intersection of the original annulus with a second annulus corresponding to the new measurement, especially when the agent moves perpendicular to the direction pointing towards the target.
The change of annular distributions into bimodal and then unimodal distributions can be rationalized
in terms of simple triangulation \cite{dobramysl2020triangulation}, see Part II. 
More complex likelihood shapes are possible in the case of prevailing exploration with large velocity in regions of small binding rate.

In this framework, we find that angular information from gradient sensing helps in discriminating the correct target direction from circle-shaped or bimodal likelihoods. In terms of entropy variations, this effect results to be greater than the Laplacian correction, which here helps only slightly by increasing the effective variation of the binding rate over spatial distances because of the convexity $\nabla^2 r=\lambda ||\mathbf{x}||^{-3}$.

\paragraph{Decision making.}
We use the average distance from the target at steady-state as our metric to evaluate the efficiency of chemotactic search. Then it is natural to introduce an \textit{exploitation} strategy, which minimizes the expected distance variation locally in time, see Eq. \eqref{d|X|} in Part II, that means to move in the direction of  $\left\langle \mathbf{x}/||\mathbf{x}|| \right\rangle$. However, for $a=0$, this strategy forces the likelihood shape to be always symmetric around the centre, and diffusion leads effectively away from the target. Having a finite $a>0$ breaks the symmetry and improves search performance, see \textsl{animation1}.

A second exploitative strategy is to go for the maximum likelihood direction, which results to have a slightly larger directional persistence, therefore enabling a more efficient comparison of concentration measurements, see \textsl{animation2}. Also here, having a finite $a>0$ improves the search performance because gradient sensing helps in removing the angular degeneracy of likelihood distributions.

The most efficient local strategy that we consider here, see Fig. (\ref{FigA1}), is the infotaxis strategy \cite{vergassola2007infotaxis, masson2009chasing, barbieri2011trajectories}, which prescribes to choose the movement direction by a maximization of the information gain locally in time. Being based on expected entropy variations only, we regard it as a purely \textit{explorative} strategy. We see that the expected entropy variation (Eq. \eqref{E dS}) is independent of the velocity, therefore we need to compute and minimize the second entropy variation with respect to the velocity direction. This gives an optimal direction
\begin{equation}\label{infotaxis a}
\mathbf{q} = \left\langle (\mathbf{\nabla} r)\,  \ln \left(\frac{\left\langle  r\right\rangle}{r}\right)  \right\rangle +  a^2 \, \mathbf{f}[p,r] ,
\end{equation}
whose precise expression of the decision correction functional $\mathbf{f}[p,r]$ is given in Part II. A realization of this dynamics is shown in \textsl{animation3}.
The impact of corrections on the infotaxis decision making is locally small, but it may be relevant macroscopically and it will be investigated in future numerical simulations.

Since we are dealing with a binding rate field, which is symmetric around the target, the integrand in the decision integral of Eq. \eqref{infotaxis a} points always in the radial direction. We can therefore interpret the magnitude of such integrand, namely $q = -p \,  ||\mathbf{\nabla} r|| \,  \ln (\langle  r\rangle/r ) +\mathcal{O}(a^2)$, as a density of \textit{decision incentive}, and simply plot it as a heatmap, see Fig. (\ref{Fig2}). In other words, we rewrite the decision integral as $\mathbf{q}= \int d\mathbf{x}\, q\, \mathbf{e^{i\theta (\mathbf{x})}}$, where $\mathbf{e^{i\theta (\mathbf{x})}}$ is the unit vector in the direction of $\mathbf{x}$. Generally, $q\lessgtr 0$ for $r\lessgtr \langle r \rangle$. We see that the infotaxis decision integral drives the agent towards regions of high likelihood, high binding rate relative to the expected $\langle  r\rangle$, and high binding rate gradient.

As benchmark to compare the performance of the three strategies considered, we study the optimal case of complete target position information. The corresponding dynamics is simply 
$d\mathbf{x} = -v (\mathbf{x}/||\mathbf{x}||) dt + \sqrt{2D}\, d\mathbf{W},$
where $\mathbf{W}$ is a standard 2D Brownian motion. Having circular symmetry, we write the SDE for the distance $||\mathbf{x}||$ using Ito's Lemma, namely 
$d\mathbf{||\mathbf{x}||} = \left( D/||\mathbf{x}|| -v\right) dt + \sqrt{2D}\, dW$,
and from stationarity $\langle\langle d||\mathbf{x}|| \rangle\rangle =0$ and Jensen's inequality, we obtain $\langle ||\mathbf{x}|| \rangle \geq D/v$. A more accurate estimate of the benchmark lower bound $b$ (which also considers the $R_\mathrm{min},R_\mathrm{max}$ boundaries) is found numerically.

\begin{center}
\begin{figure}
     \includegraphics[scale=0.58]{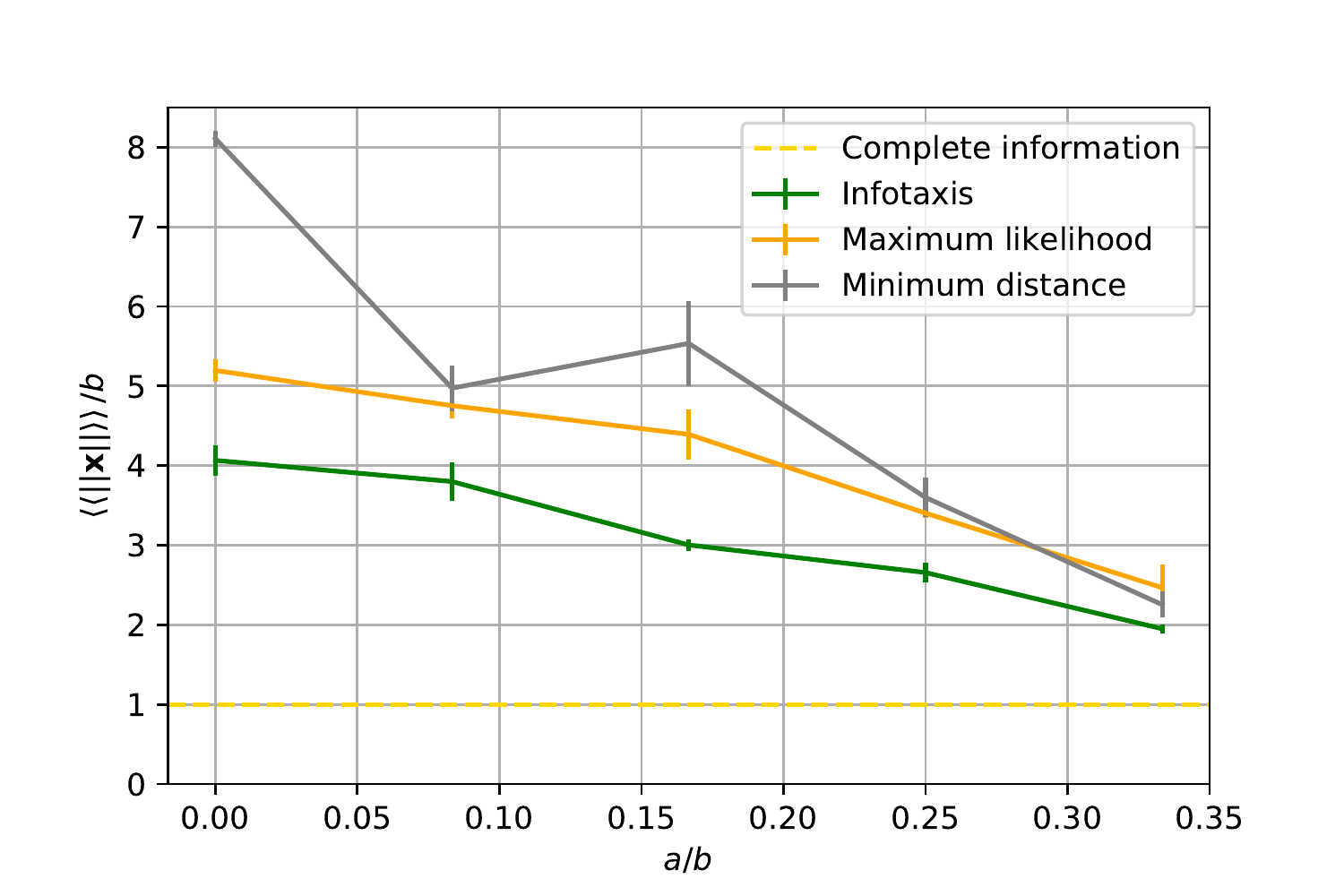}
	\caption{Search performance increases with agent's size $a$, in a steady-state dynamics with parameters $\lambda=2$, $v=0.01$, $D=2.5 \cdot 10^{-4}$, $R_\mathrm{min}=0.03$, $R_\mathrm{max}=0.87$. Blue is infotaxis, Orange is maximum likelihood decision, Green is minimum distance decision, and in Gold is the lower bound corresponding to the benchmark case of complete information, which here gives $b=0.06$.}
	\label{Fig3}
\end{figure}
\end{center}

For all three decision strategies tested, 
the mean distance from the target at steady state decreases with agent size, as expected, see Fig. (\ref{Fig3}).
For the infotaxis strategy, which generally performs best, 
this distance $\langle\langle ||\mathbf{x}|| \rangle\rangle$ was about two-fold smaller for the largest agent size tested
($a=b/3$) compared to the case of a point-like agent ($a=0$).
In fact, $\langle\langle ||\mathbf{x}||\rangle\rangle \approx 2b$ for the parameters used, i.e., comparable to the benchmark $b$ 
of complete information. 

We did not test larger agent sizes because numerical accuracy decreases with agent size, as the second order expansion will not be as accurate, and the space discretization becomes too coarse.

\paragraph{Discussion.}
In these Research Notes, we derived the information processing properties of an agent performing chemotaxis based on a likelihood map evolving with a Bayesian updating scheme, which we like to call Bayesian chemotaxis. In particular, we extend the point-like agent model of \cite{vergassola2007infotaxis} to a finite-size agent, where the knowledge of the angle of a binding event directly provides directional information.
 
Like in \cite{vergassola2007infotaxis}, we made the strong assumption that the measurement model is exact, meaning that the agent knows the binding rate dependence on distance. While this may be unrealistic for chemotaxis of bacteria, where information-processing capabilities are limited, it is certainly feasible to implement in olfactory robots \cite{chen2019odor}.

The leading order expansion terms in the expected entropy variation, which derive from having a circle of radius $a$ instead of a point-like agent, are interpreted as gradient sensing and a Laplacian correction to concentration sensing. With this expansion, we provide a theoretical framework to predict the agent size scale where the contribution from gradient sensing becomes macroscopically relevant.

In the original Ref. \cite{vergassola2007infotaxis} it was assumed that concentration sensing is the dominant contribution and effects due to the finite size of the agent can be ignored; this previous assumption is consistent with the findings reported here.
Indeed, even a Laplacian correction to concentration sensing is of the same expansion order of gradient sensing, which is instead the main ingredient of biological chemotaxis.

However, we numerically establish that the impact of gradient sensing is larger compared to the Laplacian correction if averages are taken over a steady-state chemotaxis dynamics, during which typical likelihood shapes are circles, semicircles, and bimodal shapes.
This is understood as gradient sensing, which is in some way orthogonal to concentration sensing, can remove the degeneracy in the angles likelihood, even without movement.

Finally, we observe that also in our framework infotaxis outperforms other local strategies.

\begin{acknowledgments}
AA is supported by the DFG through FR3429/3-1 to BMF; 
BMF is supported by the DFG through FR3429/4-1 (Heisenberg grant);
AA, MN, and BMF were supported through the Excellence Initiative by the German Federal and State Government (Cluster of Excellence PoL EXC-2068), and the cfaed - Center for Advancing Electronics Dresden.
\end{acknowledgments}

\begin{widetext}

\section{Part II: Derivation of the expected entropy variations equations.}

\subsection{Likelihood stochastic evolution equation for a point-like agent.}

Let us restate the setting more precisely. The search agent is equipped with a smooth likelihood distribution $p(\mathbf{x}_t)$ for the location of the target at a generic time $t$.
We choose a reference frame that is centered at and co-moving with the searcher, but keeping constant orientation relative to the environment. The agent moves with a velocity $\mathbf{v}_t$ (the target moves with velocity $-\mathbf{v}_t$ in our reference frame) having a constant modulus $||\mathbf{v}_t||=v$, and a direction angle $\theta_t$ which is continuously \textit{decided} based on the current likelihood.

Consider a noisy measurement performed during the time interval $[0,\tau)$, and expressed in integral form, $m_{\tau}=\int_0^{\tau}dm_t$. Let us denote by $\mathbf{x}_0^{t}\equiv \{ \mathbf{x}_{s} \}_0^t$ the target trajectory in the interval $[0,t)$, with $t<\tau$. The stochastic increments $dm_t$, formally defined in the limit of a small homogeneous discretization of the interval $[0,\tau)$, are Markovian and conditionally indipendent of the current likelihood given the current position, $p(dm_t|\mathbf{x}_0^{t},p(\mathbf{x}_t))=p(dm_t|\mathbf{x}_t)$.

If we instead condition the measurement probability on the whole trajectory $\mathbf{x}_0^{\tau}$, in general we have the inequality $p(dm_t|\mathbf{x}_0^{\tau},p(\mathbf{x}_t))\neq p(dm_t|\mathbf{x}_t)$, because the current measurement increment $dm_t$ affects the likelihood evolution, and on the future likelihood $p(\mathbf{x}_{t'})$ (with $t'>t$) is based the decision of the future movement direction $\theta_{t'}$, and therefore the measurement process $m_{\tau}$ has exerted a \textit{feedback}  \cite{sagawa2012nonequilibrium} to the trajectory $\mathbf{x}_0^{\tau}$.
Similarly, the stochastic increments are not conditionally independent of each other given the trajectory $\mathbf{x}_0^{\tau}\equiv \{ \mathbf{x}_t \}_0^{\tau}$, i.e., in general $p(dm_t , dm_{t'} | \mathbf{x}_0^{\tau},p(\mathbf{x}_0))\neq p(dm_t | \mathbf{x}_0^{\tau},p(\mathbf{x}_0)) p(dm_{t'} | \mathbf{x}_0^{\tau},p(\mathbf{x}_0))$.

We address these issues by imposing the decision process to happen only at discrete time instants $\{ 0,\tau,2\tau,...\}$, by delivering the outcome of the integral measurement $m_{\tau}$ to the searcher only at the end of the subinterval $t=\tau$. Taking decisions at discrete time instants means that the trajectory $\mathbf{x}_0^{\tau}$ follows a fixed protocol of directions $\theta_0^{\tau}$ decided at $t=0$, and we further assume such protocol to be continuous.  We will then take the limit $\tau\rightarrow 0$ to recover a continuous decision process. In this way there is no feedback for $t\leq \tau$, so that $p(dm_t , dm_{t'} | \mathbf{x}_0^{\tau},p(\mathbf{x}_0))= p(dm_t | \mathbf{x}_t) p(dm_{t'} | \mathbf{x}_{t'})$, and the conditional measurement probability density can be expressed in the path-integral form
\begin{equation}\label{path integral}
 p\left(m_{\tau}\big|\mathbf{x}_0^{\tau}\right)=\int d \{dm_t\}_0^{\tau}\, \delta\left( m_{\tau}-\int_0^{\tau}dm_t  \right) \,\exp\left[  \int_0^{\tau}\log p(dm_t\big| \mathbf{x}_t) \right].
\end{equation}

Before discussing the particular measurement model, note that as a result of the movement and independently of the current measurement, the likelihood is translated in the interval $[0,\tau)$ as
\begin{equation}
p(\mathbf{x}_{\tau})= p(\mathbf{x}_0) +\mathbf{\nabla} p(\mathbf{x}_0)\cdot \mathbf{v}_0 \tau +\mathcal{O}(\tau^2),
\end{equation}
which is still based on information up to $t=0$, and therefore is generally called \textit{predicted} likelihood. Indeed, it has still not been updated on the measurement outcome $m_{\tau}$. More in general, the prediction step could involve motility noise, corresponding to fluctuations of either the source or the agent itself, and we will introduce this later.

Let us consider, as a measurement model, the example of sparse molecular signals discretely detected at a rate $r(\mathbf{x})$ by an agent at the origin when the target is at position $\mathbf{x}$. For simplicity and ease of notation we assume the sensing field to be symmetric, $r(\mathbf{x})=r(-\mathbf{x})$.  The probability of observing an event in an infinitesimal interval $[t,t+dt)$ is defined as $p(dm_t=1\big| \mathbf{x}_t)=r(\mathbf{x}_t)dt$. The complementary case of not observing the event is $p(dm_t=0\big| \mathbf{x}_t)=1-r(\mathbf{x}_t)dt$. 
Now take a small finite interval $\tau$, during which the searcher moves and correspondingly its distance to the target varies.
From Eq. \eqref{path integral}, the probability of observing zero events in the interval $[0,\tau)$ is
\begin{multline}\label{path integral empty}
 p\left(m_{\tau}=0\big|\mathbf{x}_0^{\tau}\right)= \exp\left[  \int_0^{\tau}\log \left[1-r(\mathbf{x}_t)dt \right] \right]= \exp\left[  -\int_0^{\tau}r(\mathbf{x}_t)dt \right]\\
=1 -\int_0^{\tau}r(\mathbf{x}_t)dt +\mathcal{O}(\tau^2)=1 - r(\mathbf{x}_{\tau})\tau +\mathcal{O}(\tau^2),
\end{multline}
where we assumed $r(\mathbf{x})$ to be smooth in the search region (the target can still be a singularity).

From Eq. \eqref{path integral empty} we derive $p\left(m_{\tau}=0\big|\mathbf{x}_0^{\tau}\right)=p\left(m_{\tau}=0\big|\mathbf{x}_{\tau}\right)+\mathcal{O}(\tau^2)$, meaning that if we just consider $\mathbf{x}_{\tau}$ (or $\mathbf{x}_0$) and discard the details of the trajectory $\mathbf{x}_0^{\tau}$ when evaluating the measurement probability we commit an error of order $ \approx\mathbf{v}_0\cdot\mathbf{\nabla}r(\mathbf{x}_0)\tau^2=\mathcal{O}(\tau^2)$ due to the spatial dependence of the sensing rate.

For a small interval $\tau$, the probability of one single detection conditional on target position is the complementary of Eq. \eqref{path integral empty},
\begin{equation}\label{p single detection}
 p(m_{\tau}=1|\mathbf{x}_{\tau})= r(\mathbf{x}_{\tau})\tau +\mathcal{O}(\tau^2),
\end{equation}
and accordingly the case of multiple detections scales as $p(m_{\tau}\geq2)=\mathcal{O}(\tau^2)$ and is negligible compared to $p(m_{\tau}=1|\mathbf{x}_{\tau})$, which argument holds for any target position $\mathbf{x}_{\tau}$ in the limit $\tau\rightarrow 0$.

The measurement drives an update of the likelihood through Bayes rule
\begin{equation}\label{update}
 p(\mathbf{x}_{\tau}|m_{\tau})= p(\mathbf{x}_{\tau})\frac{p(m_{\tau}|\mathbf{x}_0^{\tau})}{p(m_{\tau})},
\end{equation}
where $p(m_{\tau})\equiv\int d\mathbf{x}_0^{\tau} p(\mathbf{x}_0^{\tau})p(m_{\tau}|\mathbf{x}_0^{\tau}) =\left\langle p(m_{\tau}|\mathbf{x}_0^{\tau})\right\rangle_{p(\mathbf{x}_0^{\tau})}$ is the measurement probability estimated by the searcher.
For the case $m_{\tau}=0$ it is
\begin{equation}
 p(m_{\tau}=0) \equiv \left\langle p(m_{\tau}=0|\mathbf{x}_0^{\tau})\right\rangle_{p(\mathbf{x}_0^{\tau})}=\left\langle p(m_{\tau}=0|\mathbf{x}_{\tau})\right\rangle_{p(\mathbf{x}_{\tau})}+\mathcal{O}(\tau^2)=1- \left\langle  r(\mathbf{x}_{\tau})\right\rangle \tau +\mathcal{O}(\tau^2),
\end{equation}
and the corresponding likelihood update is
\begin{equation}
 p(\mathbf{x}_{\tau}|m_{\tau}=0)= p(\mathbf{x}_{\tau})\big[1- \left[r(\mathbf{x}_{\tau})-\left\langle  r(\mathbf{x}_{\tau})\right\rangle\right]\tau\big] +\mathcal{O}(\tau^2).
\end{equation}
For the case $m_{\tau}=1$ it is
\begin{equation}
 p(m_{\tau}=1) =1-p(m_{\tau}=0)+\mathcal{O}(\tau^2)= \left\langle  r(\mathbf{x}_{\tau})\right\rangle \tau +\mathcal{O}(\tau^2),
\end{equation}
and accordingly the likelihood update reads
 \begin{equation}
 p(\mathbf{x}_{\tau}|m_{\tau}=1)= p(\mathbf{x}_{\tau})\frac{r(\mathbf{x}_{\tau}) }{\left\langle  r(\mathbf{x}_{\tau})\right\rangle  }+\mathcal{O}(\tau).
\end{equation}

The update equation is the sum of the contributions from the two possible measurement outcomes, $m_{\tau}=0$ and $m_{\tau}=1$, weighted by a Kronecker delta $\delta_{m_{\tau},1}=1-\delta_{m_{\tau},0}$ specifying the actual outcome. Considering also the prediction step we write
\begin{multline}\label{exact discrete}
p(\mathbf{x}_{\tau}|m_{\tau})- p(\mathbf{x}_0)\\= \tau\bigg\{ \mathbf{\nabla} p(\mathbf{x}_0)\cdot \mathbf{v}_0+ p(\mathbf{x}_{0})\left[- (1-\delta_{m_{\tau},1})\left[r(\mathbf{x}_{\tau})-\left\langle  r(\mathbf{x}_{\tau})\right\rangle\right] +\frac{\delta_{m_{\tau},1}}{\tau}\left( \frac{r(\mathbf{x}_{\tau}) }{\left\langle  r(\mathbf{x}_{\tau})\right\rangle  }-1 \right)\right] \bigg\} +\mathcal{O}(\tau^2)+\delta_{m_{\tau},1}\mathcal{O}(\tau)
\\= \tau\bigg\{ \mathbf{\nabla} p(\mathbf{x}_0)\cdot \mathbf{v}_0+ p(\mathbf{x}_{0})\left[- \left[r(\mathbf{x}_{\tau})-\left\langle  r(\mathbf{x}_{\tau})\right\rangle\right] +\frac{\delta_{m_{\tau},1}}{\tau}\left( \frac{r(\mathbf{x}_{\tau}) }{\left\langle  r(\mathbf{x}_{\tau})\right\rangle  }-1 \right)\right] \bigg\} +\mathcal{O}(\tau^2)+\delta_{m_{\tau},1}\mathcal{O}(\tau),
\end{multline}
where we used $p(\mathbf{x}_{\tau})=p(\mathbf{x}_{0})+\mathcal{O}(\tau)$. We neglected terms of order $\delta_{m_{\tau},1}\mathcal{O}(\tau)$ with respect to terms of order $\mathcal{O}(\tau)$ because their contribution vanishes under the integral sign. More precisely, for any given finite number of events $M\geq 0$ in a fixed finite time interval $T=n\tau$, in the limit $\tau\rightarrow 0$ ($n\rightarrow \infty$) we write
\begin{multline}
  \lim_{\tau\rightarrow 0} \tau \bigg| \sum_{j=1}^{n} p\left(\mathbf{x}_{j\tau}\big|\{m_{\tau}^{(k)}\}_{k=0}^{j-1}\right)   \delta_{m_{\tau}^{(j)},1} \left[ r(|\mathbf{x}_{j\tau}|)-\left\langle  r(|\mathbf{x}_{j\tau}|)\right\rangle \right]  \bigg| \\
\leq \lim_{\tau\rightarrow 0} \tau M \max_j p\left(\mathbf{x}_{j\tau}\big|\{m_{\tau}^{(k)}\}_{k=0}^{j-1}\right)   \max_j \big|  r(|\mathbf{x}_{j\tau}|)-\left\langle  r(|\mathbf{x}_{j\tau}|)\right\rangle  \big| \,=0,
\end{multline}
where, without loss of generality, in writing $p\left(\mathbf{x}_{j\tau}\big|\{m_{\tau}^{(k)}\}_{k=0}^{j-1}\right)$ we made the simplifying assumption that the decision process is univocally determined by the measurements, and that the likelihood density stays bounded during the interval $[0,T)$.
Please note that the number $M$ of events in a realization is independent of the discretization step $\tau$.

From Eq. \eqref{exact discrete} in the limit $\tau\rightarrow 0$ we obtain the exact stochastic evolution equation for the likelihood 
\begin{equation}
\frac{d p(\mathbf{x},t)}{p(\mathbf{x},t)} =  \left[ \mathbf{\nabla} \ln p(\mathbf{x},t)\cdot \mathbf{v} +\left\langle  r(\mathbf{x})\right\rangle - r(\mathbf{x})  \right] dt  +\left( \frac{r(\mathbf{x}) }{\left\langle  r(\mathbf{x})\right\rangle}-1 \right) dm  ,
\end{equation}
where $dm$ is the stochastic increment, which is almost surely $dm=0$, and assuming $dm=1$ only at the detection events time instants, which are happening with rate $r(\mathbf{x})$ if $\mathbf{x}$ is the actual position of the target.  Recall that expectation brakets are intended over the current likelihood $p(\mathbf{x},t)$.
Note that the multiplicative jump noise term $\frac{r(\mathbf{x}) }{\left\langle  r(\mathbf{x})\right\rangle} dm$ has to be interpreted in the Ito scheme, because of the non-anticipative nature of Bayesian updating.

To summarize, the likelihood dynamics is driven by sensing events following a Poisson statistics with rate $r(\mathbf{x})$ depending on the position of the target. The likelihood dynamics is also driven by the movement of the searcher, whose velocity direction $\theta_t$ is set by a continuous decision process.

\paragraph{Motility noise}
If the target or the searcher have a translational diffusive component of intensity $D$, ($\sqrt{2D}$ in the corresponding SDE), then the equation has a further term linear in time
\begin{equation}\label{exact}
\frac{d p}{p} =  \left[ D\frac{\mathbf{\nabla}^2 p}{p} +  \mathbf{\nabla}\ln p \cdot \mathbf{v} +\left\langle  r\right\rangle - r  \right] dt  +\left( \frac{r }{\left\langle  r\right\rangle}-1 \right) dm ,
\end{equation}
and we simplified notation, $p\equiv  p(\mathbf{x},t)$ and $r\equiv r(\mathbf{x})$.

\subsection{Introducing the decision strategies.}

An efficient search process often involves a progressive reduction of uncertainty about the target position. With this observation, one can indirectly manage the search process by setting as goal to quickly decrease the uncertainty about the target position, where uncertainty is quantified by the entropy functional of the likelihood, $S[p(\mathbf{x},t)]=-\int d\mathbf{x}\, p(\mathbf{x},t)\ln p(\mathbf{x},t)$.  This strategy is called \textit{infotaxis} \cite{vergassola2007infotaxis}, and it is entirely based on exploration. We numerically compare it with the exploitation strategies of maximum likelihood and minimum distance direction, see Part I. Even though infotaxis is defined as an entropy-based optimization, in the chemotaxis problem this strategy leads towards the target in physical space as long as the sensing rate field $r(\mathbf{x})$ is increasingly steep around the target, that means when $\mathbf{0}$ is a singularity, $d r(|\mathbf{x}|)/ d |\mathbf{x}|\rightarrow \infty$ for $|\mathbf{x}|\rightarrow 0$. 

The efficiency of a search process is generally characterized by the first passage time (FPT) distribution. Usually one wishes to optimize the FPT distribution with respect to an objective functional, that is chosen in a way to reflect the notion of efficiency in that particular search problem.
As an example, a typical goal of a search process is to minimize the expected time to find the target, i.e., the mean FPT \cite{condamin2007first}. In the presence of motility noise, which erases information from previous measurements so that the searcher may eventually get lost, a reasonable goal could be instead to maximize the probability to find the target at all, or the mean first passage time conditional to time boundaries. In all these cases, the performance quantifier depends on the parameters which specify the initial likelihood distribution. Here instead we wish to work with "natural" likelihood shapes, meaning those who arise from a steady-state search dynamics. However, for that we need to ensure the existence of a steady-state, and we do this by imposing a boundary on the agent-target distance, meaning a maximum distance $R_{max}$ which is not known by the agent and therefore not implemented in the Bayesian scheme. $R_{max}$ is the parameter of our performance quantifier.

In general, for any choice of these objectives, a global optimization is combinatorically complex because of the stochastic nature of measurements, and their nonlinear effect on the likelihood evolution and on the corresponding decisions that make the trajectory. Therefore, here we will consider only \textit{local} approximated strategies. In particular, the local infotaxis strategy consists in choosing, at each time instant, that movement direction angle $\theta$ which minimizes expected entropy variations.
Let us start by introducing the exploitation strategies.

\paragraph{Max likelihood}

The first strategy that we consider is to go in the direction of maximum likelihood.

\paragraph{Max field variation}
A second natural strategy is to maximize the expected variation of the estimated field at the agent position $\mathbf{0}$. Let us first write 
\begin{multline}
d \langle r \rangle =  \langle r \rangle_{p +d p} - \langle r \rangle
= \int d\mathbf{x} \frac{\delta  \langle r \rangle}{\delta p } d p |_{dm=0}+\left(  \langle r \rangle_{p +d p |_{dm=1}} - \langle r \rangle \right) dm
= \int d\mathbf{x} \, r  \, d p |_{dm=0}+ \left( \langle r \rangle_{p \frac{ r}{\left\langle  r \right\rangle}} -\langle r \rangle \right) dm \\
= \left\langle r  \left[ D\frac{\mathbf{\nabla}^2 p}{p} +  \mathbf{\nabla}\ln p \cdot \mathbf{v} +\left\langle  r\right\rangle - r  \right]\right\rangle\, dt+ \left( \frac{\langle r^2 \rangle}{\left\langle  r \right\rangle} -\langle r \rangle \right) dm,
\end{multline}
and then take the expectation with respect to the measurement outcome $\left\langle dm \right\rangle = \left\langle  r\right\rangle dt$, and integrate by parts to obtain
\begin{equation}\label{dr}
\left\langle d \langle r \rangle\right\rangle 
= \left( D \left\langle r  \frac{\mathbf{\nabla}^2 p}{p}  \right\rangle - \left\langle \mathbf{\nabla}r \right\rangle \cdot \mathbf{v}  \right) dt.
\end{equation}
This strategy consists in choosing the direction angle which maximizes the expected field variation, which is solved explicitly from Eq. \eqref{dr}
\begin{equation}
\mathbf{v} =\arg \max_{\mathbf{v}|v} \left\langle d \langle r \rangle\right\rangle = -v \,\frac{\left\langle \mathbf{\nabla}r \right\rangle}{\big|\big|\left\langle \mathbf{\nabla}r \right\rangle \big|\big|}.
\end{equation}

\paragraph{Min distance variation}
Similarly, consider now the expected distance variation over likelihood and measurement outcome,
\begin{equation}\label{d|X|}
\left\langle d \langle ||\mathbf{x}|| \rangle\right\rangle 
= \left( D \left\langle ||\mathbf{x}||  \frac{\mathbf{\nabla}^2 p}{p}  \right\rangle - \left\langle \frac{\mathbf{x}}{||\mathbf{x}||} \right\rangle \cdot \mathbf{v}  \right) dt.
\end{equation}
It has two terms, one for diffusion independent of strategy, and the second one which is minimized by the following minimum distance strategy,
\begin{equation}
\mathbf{v} =\arg \min_{\mathbf{v}|v} \left\langle d \langle \big|\big| \mathbf{x} \big|\big| \rangle\right\rangle = v \,\frac{\left\langle \mathbf{x}/||\mathbf{x}|| \right\rangle}{\big|\big|\left\langle  \mathbf{x}/||\mathbf{x}|| \right\rangle \big|\big|}.
\end{equation}
Note that Eq. \eqref{d|X|} is valid in general, so that for the other strategies the second term is just multiplied by the cosine of the angle difference with the minimum distance direction strategy.

\subsection{Infotaxis: entropy variations}

\paragraph{First entropy variation}

We start by writing the evolution equation for the entropy, using Eq. \eqref{exact},
\begin{multline}\label{dS}
dS[p ]\\ = S[p +d p ] -S\left[p \right]\\
= \int d\mathbf{x} \frac{\delta S[p ]}{\delta p } d p |_{dm=0}+\left( S[p +d p |_{dm=1}] -S\left[p \right] \right) dm\\
= -  \int d\mathbf{x} \, (\ln p)  \, d p |_{dm=0}+ \left( S\left[p \frac{ r}{\left\langle  r \right\rangle}\right] -S\left[p \right] \right) dm \\
= \left[ -D \left\langle  (\ln p)  \frac{\mathbf{\nabla}^2 p }{p } \right\rangle + \left\langle  (\ln p)  \left( r - \left\langle  r\right\rangle \right) \right\rangle\right]  dt
+ \left[ \ln \left\langle  r\right\rangle -\frac{\left\langle r \ln r \right\rangle}{\left\langle  r\right\rangle}  - \left\langle \left( \frac{r }{\left\langle  r\right\rangle}-1 \right) \ln p  \right\rangle  \right] dm\\
= \left[-D\left\langle  \mathbf{\nabla}^2 \ln p \right\rangle + \left\langle  (\ln p) \left( r - \left\langle  r\right\rangle \right) \right\rangle \right] dt
+ \left[ \ln \left\langle  r\right\rangle -\frac{\left\langle r \ln r \right\rangle}{\left\langle r\right\rangle}  - \left\langle \left( \frac{r }{\left\langle r\right\rangle}-1 \right) \ln p \right\rangle  \right] dm,
\end{multline}
where in the second line we neglected terms of order $dm\,d p |_{dm=0}\sim dm\,dt $ because they vanish under integral sign, then we used probability normalization, integrated by parts $\left\langle  (\ln p)  \, \mathbf{\nabla} \ln p \right\rangle=\mathbf{0}$. Note that the double integration by parts of the first term, $\left\langle ( \ln p) \frac{\mathbf{\nabla}^2 p}{p} \right\rangle=\left\langle  \mathbf{\nabla}^2 \ln p \right\rangle $, i.e., writing the entropy variation due to diffusion as the Fisher information $ -\left\langle  \mathbf{\nabla}^2 \ln p \right\rangle\geq 0$, is known as "De Bruijn identity" \cite{dembo1991information}.

Let us now take the expectation with respect to the measurement outcome $\left\langle dm \right\rangle = \left\langle  r\right\rangle dt$,
\begin{equation}\label{expected dS}
\left\langle dS \right\rangle = \left[ -D\left\langle  \mathbf{\nabla}^2 \ln p \right\rangle  +\left\langle r \,\ln \left(\frac{\langle r \rangle}{r}\right) \right\rangle \right] dt,
\end{equation}
and we identify the first term is the entropy variation as the information erasure due to diffusion, and the second term as the expected entropy variation in the update step, which we interpret as concentration sensing. 

\paragraph{Second entropy variation}
The expected entropy variation $\left\langle dS \right\rangle$ is a functional of the current likelihood, and is independent of the current velocity. The advantage of choosing a particular direction angle is found in the second expected variation, which embodies the local prediction of the change in the measurement statistics as a result of the movement,
\begin{multline}\label{dS2}
d\left\langle dS \right\rangle  = \left\langle dS \right\rangle[p+d p] -\left\langle dS \right\rangle\left[p\right]\\
= \int d\mathbf{x} \frac{\delta \left\langle dS \right\rangle[p]}{\delta p} d p|_{dm=0}+\left( \left\langle dS \right\rangle[p+d p|_{dm=1}] -\left\langle dS \right\rangle\left[p\right] \right) dm\\
= \left[ \int d\mathbf{x} \, \left( - D ( \mathbf{\nabla}^2 \ln p +\frac{\mathbf{\nabla}^2 p}{p} ) - r \left( \ln r -\ln \left\langle r\right\rangle -1  \right) \right) d p|_{dm=0}\right] dt +\left( \left\langle dS \right\rangle[p\frac{r}{\left\langle r\right\rangle}] -\left\langle dS \right\rangle\left[p\right] \right) dm\\
= \left[ \left\langle \left( \ln r - \ln \langle r \rangle \right) \, \mathbf{\nabla} r \right\rangle \cdot \mathbf{v}\,
+ \left\langle \left( - D ( \mathbf{\nabla}^2 \ln p +\frac{\mathbf{\nabla}^2 p}{p} )  - r \left( \ln r -\ln \left\langle r\right\rangle -1  \right) \right) \left( D \frac{\mathbf{\nabla}^2 p}{p} +\left\langle  r\right\rangle - r  \right) \right\rangle \right] (dt)^2 \\
- \frac{1}{\left\langle r\right\rangle} \bigg[D \left( \left\langle r \mathbf{\nabla}^2 \ln r \right\rangle + \left\langle (\left\langle r\right\rangle -r) \mathbf{\nabla}^2 \ln p \right\rangle \right) + \left\langle r^2 \ln r \right\rangle -  \left\langle r^2 \right\rangle \ln  \left\langle r^2 \right\rangle + \left\langle r^2 \right\rangle \ln  \left\langle r \right\rangle    - \left\langle r\right\rangle \left\langle r \ln r \right\rangle +\left\langle r  \right\rangle^2 \ln \left\langle r \right\rangle  \bigg] dt \,dm,
\end{multline}
where we integrated by parts $\left\langle \left( - r \left( \ln r -\ln \left\langle r\right\rangle -1  \right) \right) \, \mathbf{\nabla} \ln p   \right\rangle = \left\langle \left( \ln r - \ln \langle r \rangle \right) \, \mathbf{\nabla} r \right\rangle$, and performed vector calculus to cancel
\begin{multline}
 -\left\langle  \left( \mathbf{\nabla}^2 \ln p +\frac{\mathbf{\nabla}^2 p}{p} \right)  \, \mathbf{\nabla} \ln p   \right\rangle
= \left\langle  \mathbf{\nabla} \left( \mathbf{\nabla}^2 \ln p +\frac{\mathbf{\nabla}^2 p}{p} \right)    \right\rangle
= \left\langle  \mathbf{\nabla} \left( \mathbf{\nabla}^2 \ln p +\frac{\mathbf{\nabla}\cdot \left(p \mathbf{\nabla}\ln p \right)}{p} \right)    \right\rangle\\
= \left\langle  \mathbf{\nabla} \left( 2\mathbf{\nabla}^2 \ln p +\frac{(\mathbf{\nabla}p)\cdot \left(\mathbf{\nabla}\ln p \right)}{p} \right)    \right\rangle
= \left\langle  \mathbf{\nabla} \left( 2\mathbf{\nabla}^2 \ln p +|\mathbf{\nabla}\ln p |^2 \right)    \right\rangle\\
= 2 \left\langle  \mathbf{\nabla} \mathbf{\nabla}^2 \ln p   \right\rangle + 2\left\langle  \left(\mathbf{\nabla}\ln p \cdot  \mathbf{\nabla}\right) \mathbf{\nabla} \ln p    \right\rangle  + 2\left\langle  \mathbf{\nabla}\ln p \times  \left( \mathbf{\nabla} \times \mathbf{\nabla} \ln p  \right)  \right\rangle 
= 2 \left\langle  \mathbf{\nabla} \mathbf{\nabla}^2 \ln p   \right\rangle + 2\left\langle  \left(\mathbf{\nabla}\ln p \cdot  \mathbf{\nabla}\right) \mathbf{\nabla} \ln p    \right\rangle \\
= 2 \left\langle  \mathbf{\nabla} \mathbf{\nabla}^2 \ln p   \right\rangle - 2\left\langle  \mathbf{\nabla}^2 \mathbf{\nabla} \ln p   \right\rangle = 0.
\end{multline}

Let us now take the expectation with respect to the measurement outcome
\begin{multline}\label{expected dS2}
\left\langle d\left\langle dS \right\rangle \right\rangle \\
=  \bigg[ \left\langle \left( \ln r - \ln \langle r \rangle \right) \, \mathbf{\nabla} r \right\rangle \cdot \mathbf{v}\,
+\left\langle \left( - D ( \mathbf{\nabla}^2 \ln p +\frac{\mathbf{\nabla}^2 p}{p} )  - r \left( \ln r -\ln \left\langle r\right\rangle -1  \right) \right) \left( D \frac{\mathbf{\nabla}^2 p}{p} +\left\langle  r\right\rangle - r  \right) \right\rangle \\
 -D \left( \left\langle r \mathbf{\nabla}^2 \ln r \right\rangle +\left\langle (\left\langle r\right\rangle -r) \mathbf{\nabla}^2 \ln p \right\rangle \right) - \left\langle r^2 \ln r \right\rangle +  \left\langle r^2 \right\rangle \ln  \left\langle r^2 \right\rangle - \left\langle r^2 \right\rangle \ln  \left\langle r \right\rangle    + \left\langle r\right\rangle \left\langle r \ln r \right\rangle -\left\langle r  \right\rangle^2 \ln \left\langle r \right\rangle  \bigg]  (dt)^2.
\end{multline}

The infotaxis strategy consists in choosing the direction angle which minimizes the expected second entropy variation, which is solved explicitly from Eq. \eqref{expected dS2}
\begin{equation}\label{infotaxis}
\mathbf{v} =\arg \min_{\mathbf{v}|v} \left\langle d\left\langle dS \right\rangle \right\rangle = v \,\frac{\left\langle \left( \ln \left\langle r  \right\rangle -\ln r \right)  \mathbf{\nabla} r \right\rangle}{\big|\left\langle \left( \ln \left\langle r  \right\rangle -\ln r \right)  \mathbf{\nabla} r \right\rangle \big|}.
\end{equation}
Intuitively, that means to choose the infinitesimal displacement which maximizes the expected information gain in the next measurements, and is generally different from minimizing the expected distance variation rate.

An equivalent expression to Eq. \eqref{infotaxis} was already derived in Ref.  \cite{barbieri2011trajectories} by looking directly at the expected position-dependent information rate. Indeed the solution of Eq. \eqref{infotaxis} tells us that the infotaxis direction is the one that maximizes the expected information rate among the neighbouring positions, which is the original idea in \cite{vergassola2007infotaxis}.

\subsection{Finite-size agent: gradient sensing and Laplacian correction}

If the agent has a finite size, it can exploit receptors in different locations on its body to measure gradients \cite{novak2021bayesian}. In this way, given that the field shape $r(\mathbf{x})$ is known, in a sufficiently long measurement time the agent is able to exactly locate the target (if the local field shape univocally determines the target position), even without moving.

Let us assume that detection events happen only on the surface of the agent, which in 2D is a circle of radius $a$. If we neglect the effect of the agent body and movement on the sensing field, the inference problem is equivalent to the inference made by a density of agents distributed on the circle $a\{\mathbf{e^{i\theta}}\}_{\theta=[0,2\pi]}$, where $\mathbf{e^{i\theta}}$ is the unit vector corresponding to the angle $\theta$.
Accordingly, the event probability density for unit angle in a small time interval $\tau$ is $p(\mathbf{m}_{\tau}= \mathbf{e^{i\theta}}|\mathbf{x})=\frac{\tau}{2\pi} r(\mathbf{x}-a\mathbf{e^{i\theta}})+\mathcal{O}(\tau^2)$. Note that, with this normalization, in the limit $a\rightarrow 0$ we recover the case of point-like agent discussed in previous sections. Also note that the measurement $\mathbf{m}_{\tau}$ is now vector-valued, assuming $\mathbf{0}$ almost surely (for $\tau\rightarrow 0$), and $\mathbf{e^{i\theta}}$ in the case of a binding event at angle $\theta$. 

Here we consider gradient sensing in the second order expansion of the sensing field in the agent dimension $a$,
\begin{equation}\label{a expansion}
r(\mathbf{x}-a\mathbf{e^{i\theta}}) = r(\mathbf{x}) -a\,\mathbf{e^{i\theta}} \cdot \mathbf{\nabla} r(\mathbf{x}) +\frac{1}{2} a^2 \mathbf{e^{i\theta}}\cdot\left( \mathbf{H} (\mathbf{x}) \,\mathbf{e^{i\theta}} \right) \,+ \mathcal{O}(a^3),
\end{equation}
where $\mathbf{H}(\mathbf{x})\equiv \mathbf{\nabla} \otimes \mathbf{\nabla} r(\mathbf{x})$ is the Hessian matrix of the sensing field at $\mathbf{x}$.
Note that this expansion is not well-defined at the singularities of $r(\mathbf{x})$, therefore we assume that the likelihood vanishes enough fast around singularities, or that $r(\mathbf{x})$ simply does not have singularities.

The likelihood update in the case of one event at angle $\theta$ reads
 \begin{multline}
 p\left(\mathbf{x}_{\tau}\big|\mathbf{m}_{\tau}= \mathbf{e^{i\theta}}\right)=
 p(\mathbf{x}_{\tau})\frac{p\left(\mathbf{m}_{\tau}= \mathbf{e^{i\theta}}\big|\mathbf{x}_{\tau}\right)}{p\left(\mathbf{m}_{\tau}= \mathbf{e^{i\theta}}\right)}+\mathcal{O}(\tau)
=  p(\mathbf{x}_{\tau})\frac{r(\mathbf{x}_{\tau}-a\mathbf{e^{i\theta}}) }{\left\langle  r(\mathbf{x}_{\tau}-a\mathbf{e^{i\theta}})\right\rangle  }+\mathcal{O}(\tau)\\
=  p(\mathbf{x}_{\tau}) \frac{r(\mathbf{x}_{\tau}) }{\left\langle  r(\mathbf{x}_{\tau})\right\rangle }\bigg[1+a \mathbf{e^{i\theta}}\cdot \left( \frac{\left\langle \mathbf{\nabla} r(\mathbf{x}_{\tau})\right\rangle}{\langle r(\mathbf{x}_{\tau})\rangle} - \frac{\mathbf{\nabla} r(\mathbf{x}_{\tau})}{r(\mathbf{x}_{\tau})} \right)
 +\frac{1}{2} a^2   \mathbf{e^{i\theta}}\cdot\left(\left[ \frac{\mathbf{H} (\mathbf{x}_{\tau})}{r(\mathbf{x}_{\tau})} -\frac{\left\langle \mathbf{H} (\mathbf{x}_{\tau}) \right\rangle}{\langle r(\mathbf{x}_{\tau})\rangle}  \right]   \,\mathbf{e^{i\theta}} \right)  \\
+a^2 \left( \mathbf{e^{i\theta}}\cdot \frac{\left\langle \mathbf{\nabla} r(\mathbf{x}_{\tau})\right\rangle}{\langle r(\mathbf{x}_{\tau})\rangle} \right)  \left( \mathbf{e^{i\theta}}\cdot \left[\frac{\left\langle \mathbf{\nabla} r(\mathbf{x}_{\tau})\right\rangle}{\langle r(\mathbf{x}_{\tau})\rangle}-\frac{\mathbf{\nabla} r(\mathbf{x}_{\tau})}{ r(\mathbf{x}_{\tau})} \right] \right) \bigg] \, +\mathcal{O}(a^3) \,+\mathcal{O}(\tau),
\end{multline}
where we used the expansion of Eq. \eqref{a expansion}. The likelihood update in the case of no events reads
\begin{multline}
 p\left(\mathbf{x}_{\tau}\big|\mathbf{m}_{\tau}= \mathbf{0}\right)=\\
 p(\mathbf{x}_{\tau})\frac{p\left(\mathbf{m}_{\tau}= \mathbf{0}\big|\mathbf{x}_{\tau}\right)}{p\left(\mathbf{m}_{\tau}= \mathbf{0}\right)}+\mathcal{O}(\tau^2)\,\,
=p(\mathbf{x}_{\tau})\frac{1-\int_0^{2\pi} d\theta\, p\left(\mathbf{m}_{\tau}= \mathbf{e^{i\theta}}\big|\mathbf{x}_{\tau}\right)}{1-\int_0^{2\pi} d\theta\, p\left(\mathbf{m}_{\tau}= \mathbf{e^{i\theta}}\right)}+\mathcal{O}(\tau^2)\\
=  p(\mathbf{x}_{\tau})\left[1+\frac{\tau}{2\pi}\int_0^{2\pi}d\theta\,\left[\left\langle  r(\mathbf{x}_{\tau}-a\mathbf{e^{i\theta}})\right\rangle- r(\mathbf{x}_{\tau}-a\mathbf{e^{i\theta}})\right] \right] +\mathcal{O}(\tau^2)\\
=  p(\mathbf{x}_{\tau})\left[1+\tau \left(\left\langle  r(\mathbf{x}_{\tau})\right\rangle- r(\mathbf{x}_{\tau}) +\frac{a^2}{4\pi}\int_0^{2\pi}d\theta\,\mathbf{e^{i\theta}}\cdot\left(\left[\left\langle \mathbf{H} (\mathbf{x}_{\tau}) \right\rangle- \mathbf{H} (\mathbf{x}_{\tau}) \right]\,\mathbf{e^{i\theta}} \right) \right) \right] +\mathcal{O}(\tau a^3) +\mathcal{O}(\tau^2)\\
=  p(\mathbf{x}_{\tau})\left[1+\tau \left(\left\langle  r(\mathbf{x}_{\tau})\right\rangle- r(\mathbf{x}_{\tau}) +\frac{a^2}{4}\left[\left\langle \nabla^2 r(\mathbf{x}_{\tau}) \right\rangle- \nabla^2 r(\mathbf{x}_{\tau}) \right] \right) \right] +\mathcal{O}(\tau a^3) +\mathcal{O}(\tau^2),
\end{multline}
where the linear order in $a$ vanishes because $\int_0^{2\pi}d\theta\, \mathbf{e^{i\theta}}=\mathbf{0}$, and we used $\int_0^{2\pi}d\theta\, \cos^2(\theta) = \pi$ and $\int_0^{2\pi}d\theta\, \cos(\theta)\sin(\theta) = 0$. Also note that $(\mathbf{H} )_{11}+(\mathbf{H} )_{22} =  \nabla^2 r$.

As usual, taking the limit $\tau\rightarrow 0$ we obtain the likelihood evolution equation in the Ito scheme,
\begin{multline}\label{gradient sensing}
\frac{d p}{p} =  \left[ D\frac{\mathbf{\nabla}^2 p}{p} +  \mathbf{\nabla}\ln p \cdot \mathbf{v} +\left\langle  r\right\rangle - r +\frac{a^2}{4}\left(\left\langle \nabla^2 r \right\rangle- \nabla^2 r \right) \right] dt \\
 +\left( \frac{r }{\left\langle  r\right\rangle}-1 \right) \,||d\mathbf{m}||^2 + a \,\frac{r }{\left\langle  r\right\rangle} \left( \frac{\left\langle \mathbf{\nabla} r\right\rangle}{\left\langle  r\right\rangle}-\frac{\mathbf{\nabla} r }{r} \right) \cdot d\mathbf{m}\\
 + \frac{1}{2} a^2 \frac{r }{\left\langle  r\right\rangle} \left[   d\mathbf{m}\cdot\left(\left[ \frac{\mathbf{H} }{r} -\frac{\left\langle \mathbf{H} \right\rangle}{\langle r\rangle}  \right]   \,d\mathbf{m} \right)  + 2 \left( d\mathbf{m}\cdot \frac{\left\langle \mathbf{\nabla} r\right\rangle}{\langle r\rangle} \right) \left( d\mathbf{m}\cdot \left[\frac{\left\langle \mathbf{\nabla} r\right\rangle}{\langle r\rangle} -\frac{ \mathbf{\nabla} r}{ r}\right] \right) \right],
\end{multline}
where $||d\mathbf{m}||^2=d\mathbf{m}\cdot d\mathbf{m}$ denotes the squared modulus. In the limit of $a\rightarrow 0$ this agrees with the dimensionless agent case (Eq. \eqref{exact}) with $dm$ replaced by $||d\mathbf{m}||^2$.

The term $a^2\left(\left\langle \nabla^2 r \right\rangle- \nabla^2 r \right)$ is an effect of the sensing field's convexity, which is taken into account by the search agent. Indeed, in the absence of events, those target locations corresponding to high convexity of the sensing field have a negative correction in the likelihood evolution, because the convexity increases the rate of events proportionally to the squared radius $a^2$.

\subsection{Infotaxis: entropy variations for a finite-size agent}

The decision step is unaffected for the "Max likelihood", "Max field variation", and "Min distance variation" strategies, unless the objectives are redefined as properties on the surface of the agent, but we are not exploring this.
Here we are interested in how the decision changes when the effect of gradient sensing measurements, i.e. to consider where on the agent's body the binding events happen, is taken into account. Intuitively, the agent would move towards regions of expected high sensing field gradient and convexity to exploit its gradient sensing capabilities.

\paragraph{First entropy variation}
To study the effect of gradient sensing on the infotaxis strategy, let us first write the entropy variation, using Eq. \eqref{gradient sensing},
\begin{multline}\label{dS}
dS[p ]\\ = S[p +d p ] -S\left[p \right]\\
= \int d\mathbf{x} \frac{\delta S[p ]}{\delta p } d p |_{d\mathbf{m}=\mathbf{0}}\,+\left( S[p +d p |_{d\mathbf{m}}] -S\left[p \right] \right) \, ||d\mathbf{m}||^2\\
= -  \int d\mathbf{x} \, (\ln p)  \, d p |_{d\mathbf{m}=\mathbf{0}}\\+ \bigg( S\left[p \frac{ r}{\left\langle  r \right\rangle}\left\{ 1+ a \, \left( \frac{\left\langle \mathbf{\nabla} r\right\rangle}{\left\langle  r\right\rangle}-\frac{\mathbf{\nabla} r }{r} \right) \cdot d\mathbf{m} + \frac{1}{2} a^2  \left[   d\mathbf{m}\cdot\left(\left[ \frac{\mathbf{H} }{r} -\frac{\left\langle \mathbf{H} \right\rangle}{\langle r\rangle}  \right]   \,d\mathbf{m} \right)  + 2 \left( d\mathbf{m}\cdot \frac{\left\langle \mathbf{\nabla} r\right\rangle}{\langle r\rangle} \right) \left( d\mathbf{m}\cdot \left[\frac{\left\langle \mathbf{\nabla} r\right\rangle}{\langle r\rangle} -\frac{ \mathbf{\nabla} r}{ r}\right] \right) \right] \right\}\right]\\
 -S\left[p \right] \bigg) \, ||d\mathbf{m}||^2 \\
= \left[-D\left\langle  \mathbf{\nabla}^2 \ln p \right\rangle + \left\langle  (\ln p) \left( r - \left\langle  r\right\rangle \right) \right\rangle + \frac{a^2}{4} \left\langle  (\ln p) \left(\nabla^2 r - \left\langle \nabla^2 r \right\rangle \right) \right\rangle \right] dt\\
+ \left[ \ln \left\langle  r\right\rangle -\frac{\left\langle r \ln r \right\rangle}{\left\langle r\right\rangle}  - \left\langle \left( \frac{r }{\left\langle r\right\rangle}-1 \right) \ln p \right\rangle  \right] \, ||d\mathbf{m}||^2
+a \left\langle \frac{r}{\left\langle r \right\rangle} \left( 1+\ln\left( p \frac{r}{\left\langle r \right\rangle}  \right) \right) \left( \frac{\mathbf{\nabla} r }{r} -\frac{\left\langle \mathbf{\nabla} r\right\rangle}{\left\langle  r\right\rangle} \right) \right\rangle \cdot d\mathbf{m}  \\
-\frac{1}{2} a^2 \left\langle \frac{r}{\left\langle r \right\rangle}  \left( 1+\ln\left( p \frac{r}{\left\langle r \right\rangle}  \right) \right)  \left[   d\mathbf{m}\cdot\left(\left[ \frac{\mathbf{H} }{r} -\frac{\left\langle \mathbf{H} \right\rangle}{\langle r\rangle}  \right]   \,d\mathbf{m} \right)  + 2 \left( d\mathbf{m}\cdot \frac{\left\langle \mathbf{\nabla} r\right\rangle}{\langle r\rangle} \right) \left( d\mathbf{m}\cdot \left[\frac{\left\langle \mathbf{\nabla} r\right\rangle}{\langle r\rangle} -\frac{ \mathbf{\nabla} r}{ r}\right] \right) \right]  \right\rangle \\
-\frac{1}{2} a^2 \left\langle \frac{r}{\left\langle r \right\rangle}   \left[ \left( \frac{\left\langle \mathbf{\nabla} r\right\rangle}{\left\langle  r\right\rangle}-\frac{\mathbf{\nabla} r }{r} \right) \cdot d\mathbf{m} \right]^2 \right\rangle.
\end{multline}

Let us evaluate the expectations of the terms dependent on the stochastic measurement increment $d\mathbf{m}$ entering Eq. \eqref{dS}, with respect to the measurement probability $p(d\mathbf{m})=\left\langle p(d\mathbf{m}\big|\mathbf{x})\right\rangle$. The expected measurement increment is
\begin{multline}
\left\langle d\mathbf{m} \right\rangle \,
=\int d\mathbf{x} \,p(\mathbf{x}) \int_0^{2\pi}d\theta\, p(d\mathbf{m}= \mathbf{e^{i\theta}}\big|\mathbf{x})\,\mathbf{e^{i\theta}}\,
= \int_0^{2\pi}d\theta\, \left\langle p(d\mathbf{m}= \mathbf{e^{i\theta}}\big|\mathbf{x})\right\rangle\,\mathbf{e^{i\theta}}\\
= -a\,\frac{dt}{2\pi} \int_0^{2\pi}d\theta\, \left( \mathbf{e^{i\theta}} \cdot \left\langle \mathbf{\nabla} r \right\rangle \right)\,\mathbf{e^{i\theta}}\,+\mathcal{O}(dt \,a^2)\,
= -a\,\frac{dt}{2\pi} \left\langle \mathbf{\nabla} r \right\rangle \int_0^{\pi}d\theta\, 2\cos^2(\theta)\,+\mathcal{O}(dt \,a^2)\\
= -a\,\frac{dt}{2} \left\langle \mathbf{\nabla} r\right\rangle\,+\mathcal{O}(dt \,a^2).
\end{multline}
The expected squared norm of the measurement increment is
\begin{multline}\label{r bar}
\left\langle ||d\mathbf{m}||^2 \right\rangle \,
=\int d\mathbf{x} \,p(\mathbf{x}) \int_0^{2\pi}d\theta\, p(d\mathbf{m}= \mathbf{e^{i\theta}}\big|\mathbf{x})\,
= \int_0^{2\pi}d\theta\, \left\langle p(d\mathbf{m}= \mathbf{e^{i\theta}}\big|\mathbf{x})\right\rangle
= \left(\left\langle r \right\rangle + \frac{a^2}{4} \left\langle \nabla^2 r \right\rangle \right) dt +\mathcal{O}(dt \,a^3).
\end{multline}
Let us evaluate the squared dot product
\begin{multline}
\left\langle \left\langle \frac{r}{\left\langle r \right\rangle}   \left[ \left( \frac{\left\langle \mathbf{\nabla} r\right\rangle}{\left\langle  r\right\rangle}-\frac{\mathbf{\nabla} r }{r} \right) \cdot d\mathbf{m} \right]^2 \right\rangle \right\rangle
\equiv \left\langle \left\langle \frac{r}{\left\langle r \right\rangle}   \left[ \left( \frac{\left\langle \mathbf{\nabla} r\right\rangle}{\left\langle  r\right\rangle}-\frac{\mathbf{\nabla} r }{r} \right) \cdot d\mathbf{m} \right]^2 \right\rangle_{p(\mathbf{x})} \right\rangle_{p(d\mathbf{m})} \\
= \left\langle \frac{r}{\left\langle r \right\rangle}  \left\langle \left[ \left( \frac{\left\langle \mathbf{\nabla} r\right\rangle}{\left\langle  r\right\rangle}-\frac{\mathbf{\nabla} r }{r} \right) \cdot d\mathbf{m} \right]^2 \right\rangle_{p(d\mathbf{m})} \right\rangle_{p(\mathbf{x})}\,
= \left\langle \frac{r}{\left\langle r \right\rangle} \int_0^{2\pi} d\theta\,  p(d\mathbf{m}= \mathbf{e^{i\theta}}) \left[ \left( \frac{\left\langle \mathbf{\nabla} r\right\rangle}{\left\langle  r\right\rangle}-\frac{\mathbf{\nabla} r }{r} \right) \cdot  \mathbf{e^{i\theta}} \right]^2  \right\rangle\\
= \left\langle  r \frac{1}{2\pi} \int_0^{2\pi} d\theta\,  \left[ \left( \frac{\left\langle \mathbf{\nabla} r\right\rangle}{\left\langle  r\right\rangle}-\frac{\mathbf{\nabla} r }{r} \right) \cdot  \mathbf{e^{i\theta}} \right]^2  \right\rangle dt +\mathcal{O}(a\,dt)\,
= \frac{1}{2} \left\langle r \, \bigg|\bigg|  \frac{\left\langle \mathbf{\nabla} r\right\rangle}{\left\langle  r\right\rangle}-\frac{\mathbf{\nabla} r }{r} \bigg|\bigg| ^2  \right\rangle\, dt +\mathcal{O}(a\,dt),
\end{multline}
and similarly
\begin{multline}
2 \left\langle \left\langle  \frac{r}{\left\langle r \right\rangle}  \left( 1+\ln\left( p \frac{r}{\left\langle r \right\rangle}  \right) \right) \left( d\mathbf{m}\cdot \frac{\left\langle \mathbf{\nabla} r\right\rangle}{\langle r\rangle} \right) \left( d\mathbf{m}\cdot \left[\frac{\left\langle \mathbf{\nabla} r\right\rangle}{\langle r\rangle} -\frac{ \mathbf{\nabla} r}{ r}\right] \right)   \right\rangle \right\rangle\\
 = \left\langle r  \left(\ln\left( p \frac{r}{\left\langle r \right\rangle}  \right)\right) \frac{\left\langle \mathbf{\nabla} r\right\rangle}{\langle r\rangle} \cdot  \left[\frac{\left\langle \mathbf{\nabla} r\right\rangle}{\langle r\rangle} -\frac{ \mathbf{\nabla} r}{ r}\right] \right\rangle \,dt +\mathcal{O}(a\,dt).
\end{multline}
The last term to evaluate is
\begin{multline}
 \left\langle  \left\langle \frac{r}{\left\langle r \right\rangle}  \left( 1+\ln\left( p \frac{r}{\left\langle r \right\rangle}  \right) \right)  \,  d\mathbf{m}\cdot\left(\left[ \frac{\mathbf{H} }{r} -\frac{\left\langle \mathbf{H} \right\rangle}{\langle r\rangle}  \right]   \,d\mathbf{m} \right)  \right\rangle \right\rangle
=\frac{1}{2}\left\langle  \left(\ln\left( p \frac{r}{\left\langle r \right\rangle}  \right)\right)  \,  \left( \nabla^2 r -r \frac{\left\langle \nabla^2 r \right\rangle}{\langle r\rangle}  \right)    \right\rangle\,dt +\mathcal{O}(a\,dt).
\end{multline}

The expected entropy variation, up to second order in $a$, is then
\begin{equation}
\frac{\left\langle dS \right\rangle}{dt} 
= -D\left\langle  \mathbf{\nabla}^2 \ln p \right\rangle  
+  \left\langle \left(r +\frac{a^2}{4}\nabla^2 r \right) \ln \left(\frac{\left\langle  r\right\rangle}{r}\right) \right\rangle 
+\frac{a^2}{4}  \left( \frac{|| \langle\mathbf{\nabla} r \rangle ||^2}{\langle r \rangle}-\left\langle \frac{|| \mathbf{\nabla} r ||^2}{r}  \right\rangle  \right) \, +\mathcal{O}(a^3),
\end{equation}
which is Eq. \eqref{E dS} of Part I. As already stated in Part I, the terms in equation \eqref{E dS} are interpreted as follows:\\
\begin{itemize}
\item Order $0$:
\begin{itemize}
\item Information erasure (diffusion): $-D\left\langle  \mathbf{\nabla}^2 \ln p \right\rangle \geq 0$.
\item Concentration sensing: $  \left\langle r\, \ln \left(\frac{\left\langle  r\right\rangle}{r}\right) \right\rangle  \leq 0  $.
\end{itemize}
\item Order $a^2$:
\begin{itemize}
\item Gradient sensing: $ \frac{|| \langle\mathbf{\nabla} r \rangle ||^2}{\langle r \rangle}-\left\langle \frac{|| \mathbf{\nabla} r ||^2}{r}  \right\rangle \leq 0 $.
\item Laplacian correction to concentration sensing: $ \left\langle (\nabla^2 r ) \,\ln \left(\frac{\left\langle  r\right\rangle}{r}\right) \right\rangle   $.
\end{itemize}
\end{itemize}
We observe that the information gain from gradient sensing and the Laplacian correction are of same order in Bayesian chemotaxis. This derives from the fact that the agent knows exactly the binding rate field, so that concentration sensing dominates the Bayesian updating.

\paragraph{Second entropy variation}
Let us now evaluate the second entropy variation,
\begin{multline}\label{dS2 a}
d\left\langle dS \right\rangle  = \left\langle dS \right\rangle[p+d p] -\left\langle dS \right\rangle\left[p\right]\\
=  \left\langle \frac{\delta \left\langle dS \right\rangle[p]}{\delta p} \mathbf{\nabla} \ln p \right\rangle  \cdot  \mathbf{v} \,(dt)^2 + .........\\
=  \left\langle \left( - D ( \mathbf{\nabla}^2 \ln p +\frac{\mathbf{\nabla}^2 p}{p} ) - r \left( \ln r -\ln \left\langle r\right\rangle -1  \right) \right)\mathbf{\nabla} \ln p \right\rangle  \cdot  \mathbf{v} \,(dt)^2 \\
+ \frac{a^2}{4}  \left\langle \left( \left(\nabla^2 r\right) \left( \ln \left\langle  r\right\rangle -\ln r \right)  +\frac{\left\langle \nabla^2 r\right\rangle}{\left\langle  r\right\rangle} r
 -\frac{|| \mathbf{\nabla} r ||^2}{r}  - \frac{|| \langle\mathbf{\nabla} r \rangle ||^2}{\langle r \rangle^2} r + 2 \frac{ \langle\mathbf{\nabla} r \rangle }{\langle r \rangle} \cdot  \mathbf{\nabla} r            \right) \mathbf{\nabla} \ln p \right\rangle  \cdot  \mathbf{v} \,(dt)^2  +  ..........\\
=  \left\langle \left( \ln r - \ln \langle r \rangle \right) \, \mathbf{\nabla} r \right\rangle \cdot \mathbf{v} \,(dt)^2 \\
- \frac{a^2}{4} \bigg[ - \left\langle ( \mathbf{\nabla} \nabla^2 r)  \ln r  \right\rangle  +\left\langle  \mathbf{\nabla} \nabla^2 r \right\rangle     \ln \left\langle r  \right\rangle - \left\langle\frac{(\nabla^2 r) \mathbf{\nabla} r}{r} \right\rangle
+ \frac{\langle \nabla^2 r\rangle \langle \mathbf{\nabla} r \rangle}{ \langle r \rangle} +2 \frac{\langle \mathbf{\nabla} r \rangle}{\langle r \rangle} \cdot \langle \mathbf{\nabla}\mathbf{\nabla} r \rangle -2 \left\langle\frac{ \mathbf{\nabla} r }{ r} \cdot \mathbf{\nabla}\mathbf{\nabla} r \right\rangle\\
-\frac{||\langle \mathbf{\nabla} r \rangle||^2\, \langle \mathbf{\nabla} r \rangle}{\langle r \rangle^2} + \left\langle \frac{|| \mathbf{\nabla} r ||^2\,  \mathbf{\nabla} r  }{  r^2} \right \rangle
 \bigg] \cdot \mathbf{v} \,(dt)^2  + ..........,
\end{multline}
where other terms $..........$ independent of $\mathbf{v}$ were not considered.

The infotaxis strategy consists in choosing the direction angle which minimizes the expected second entropy variation, which is solved explicitly from Eq. \eqref{dS2 a},
\begin{equation}
\mathbf{v} =\arg \min_{\mathbf{v}|v} \left\langle d\left\langle dS \right\rangle \right\rangle = v \,\frac{\mathbf{q}}{||\mathbf{q} ||},
\end{equation}
with
\begin{multline}
  \mathbf{q} = \left\langle (\mathbf{\nabla} r) \left( \ln \langle r \rangle - \ln r \right) \right\rangle 
+ \frac{a^2}{4} \bigg[ - \left\langle ( \mathbf{\nabla} \nabla^2 r)  \ln r  \right\rangle  +\left\langle  \mathbf{\nabla} \nabla^2 r \right\rangle     \ln \left\langle r  \right\rangle - \left\langle\frac{(\nabla^2 r) \mathbf{\nabla} r}{r} \right\rangle
+ \frac{\langle \nabla^2 r\rangle \langle \mathbf{\nabla} r \rangle}{ \langle r \rangle} \\
+2 \frac{\langle \mathbf{\nabla} r \rangle}{\langle r \rangle} \cdot \langle \mathbf{\nabla}\mathbf{\nabla} r \rangle -2 \left\langle\frac{ \mathbf{\nabla} r }{ r} \cdot \mathbf{\nabla}\mathbf{\nabla} r \right\rangle
-\frac{||\langle \mathbf{\nabla} r \rangle||^2\, \langle \mathbf{\nabla} r \rangle}{\langle r \rangle^2} + \left\langle \frac{|| \mathbf{\nabla} r ||^2\,  \mathbf{\nabla} r  }{  r^2} \right \rangle
 \bigg] .
\end{multline}

\begin{center}
\begin{figure}
     \includegraphics[scale=0.48, trim = {1cm 10 1 10},clip]{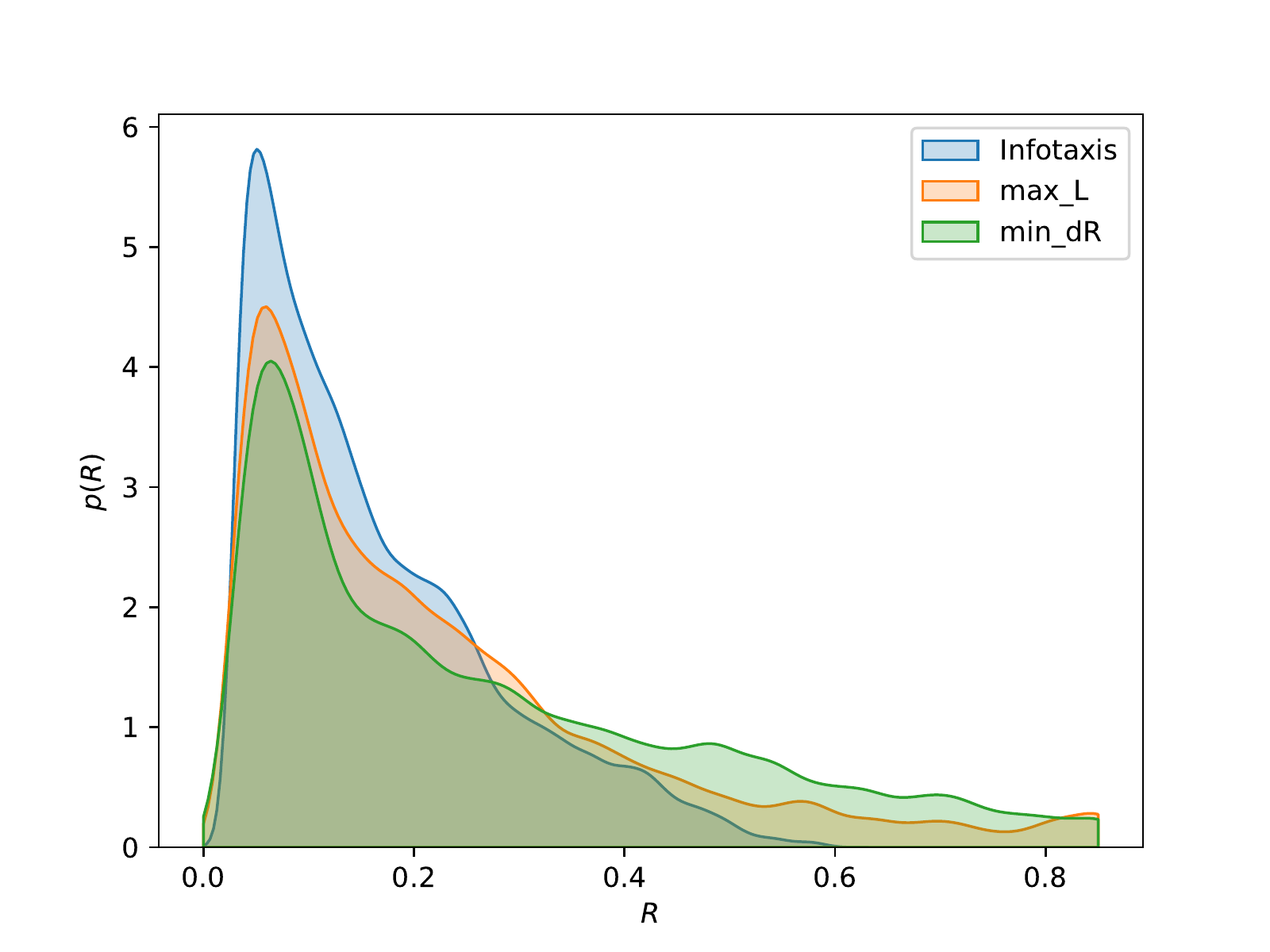}
	\caption{Infotaxis outperforms other strategies in a steady-state dynamics with parameters $a=0.01$, $\lambda=2$, $v=0.01$, $D=2.5 \cdot 10^{-4}$, $R_\mathrm{min}=0.03$, $R_\mathrm{max}=0.87$.}
	\label{FigA1}
\end{figure}
\end{center}

\subsection{Triangulation}

We can get some intuition about the likelihood shapes arising in the dynamics by considering a toy example of triangulation, see Fig. (\ref{FigA2}), where circle and bimodal likelihood shapes appear. The chemotaxis dynamics for $a=0$ can be though of as a generalization of this toy example, where the finite velocity and finite measurement precision give rise to more complex likelihood shapes. Having $a>0$ further polarizes measurements by gradient sensing. 
\begin{center}
\begin{figure}
     \includegraphics[scale=0.27, trim = {1cm 1 1 1},clip]{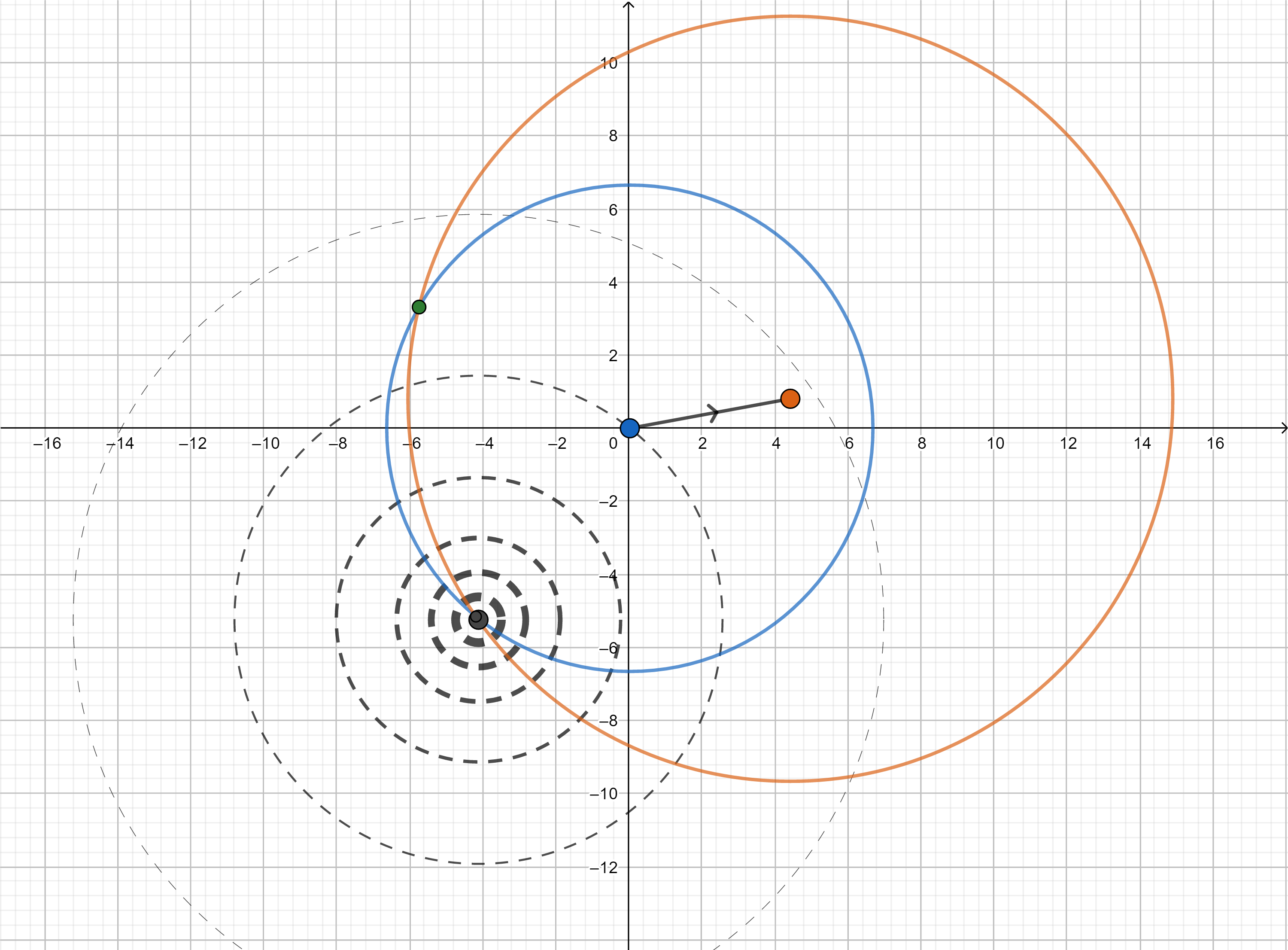}
	\caption{Triangulation example.  It consists of an agent who performs noiseless distance measurements, and moves by discrete steps. The agent starts at time $t=0$ at the origin (Blue point), and performs a first measurement, so that its likelihood is a perfect circle of radius equivalent to the target distance (Blue circle). Note the target is the black point. Then the agent moves taking a finite step in a random direction up to time $t=1$ (Orange point), and then performs a second measurement, which means a second circle (Orange circle). The two circles will intersect in two points (Black true target and a second Green intersection point), giving rise to a bimodal likelihood distribution. With a further step and measurement (not shown), the agent resolves all the uncertainty and locates the target.}
	\label{FigA2}
\end{figure}
\end{center}

\begin{center}
\begin{figure}
     \includegraphics[scale=0.48, trim = {0.6cm 50 1 50}, clip]{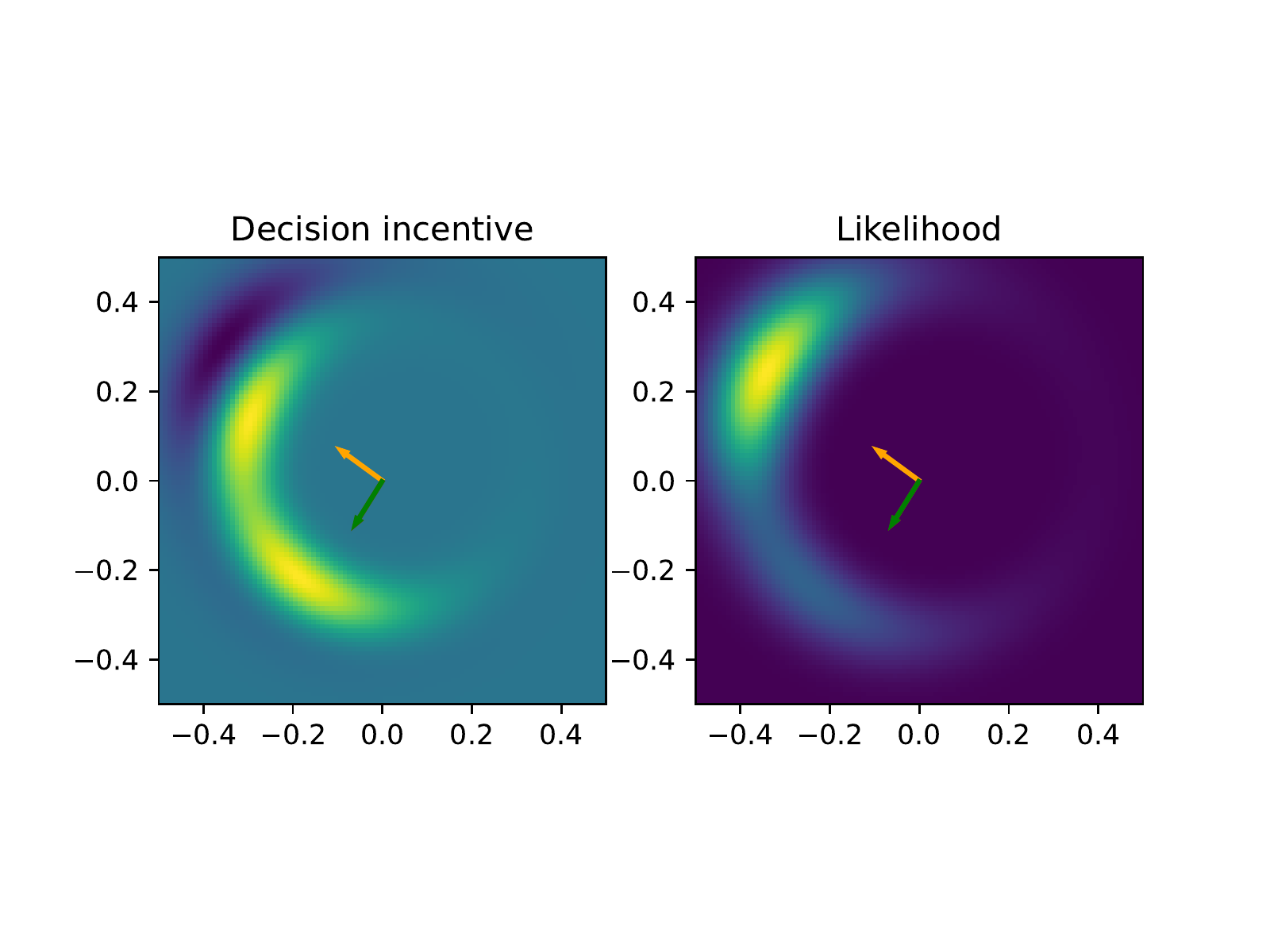}
     \includegraphics[scale=0.48, trim = {0.6cm 50 1 50}, clip]{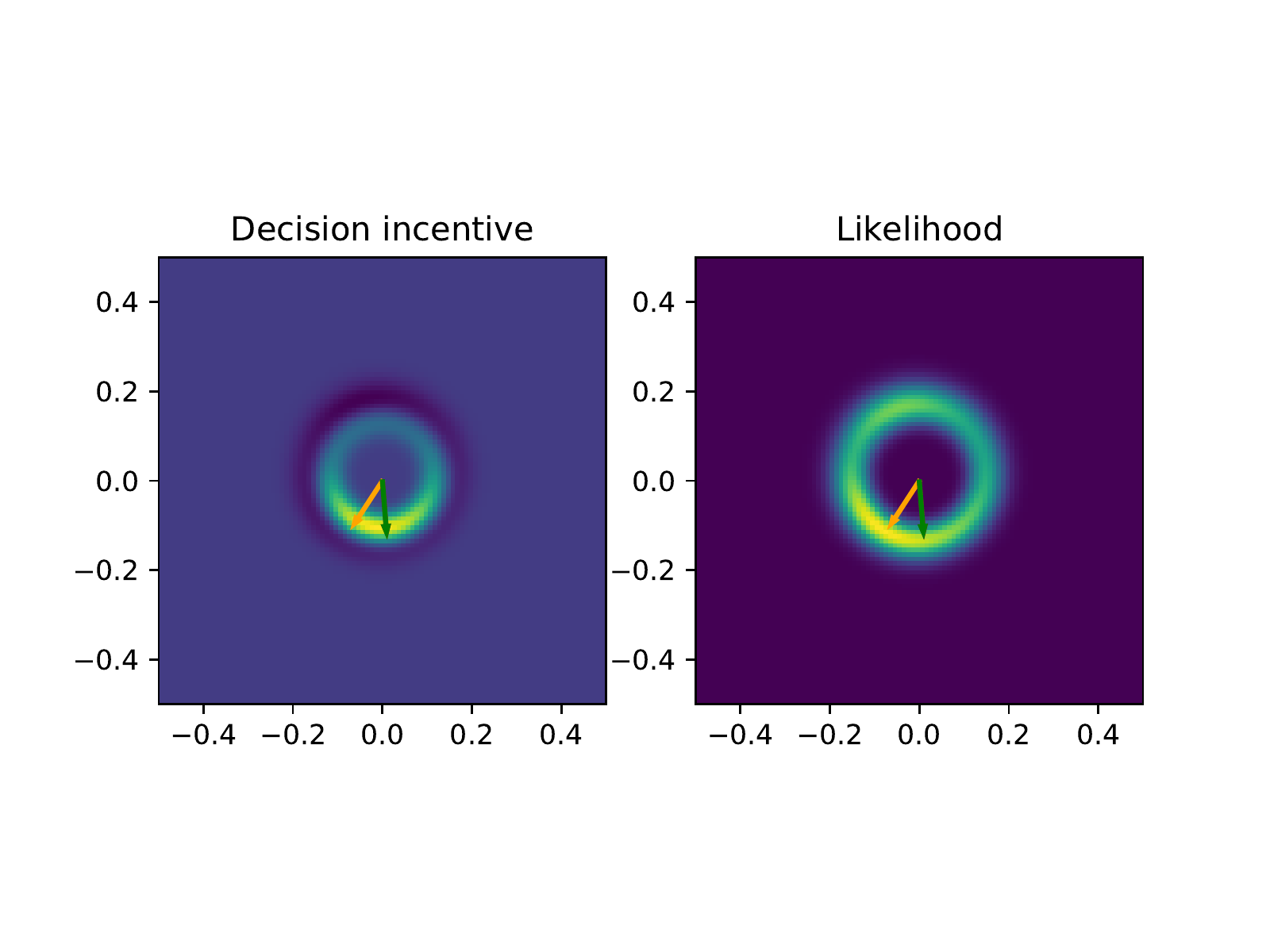}
     \includegraphics[scale=0.48, trim = {0.6cm 50 1 50}, clip]{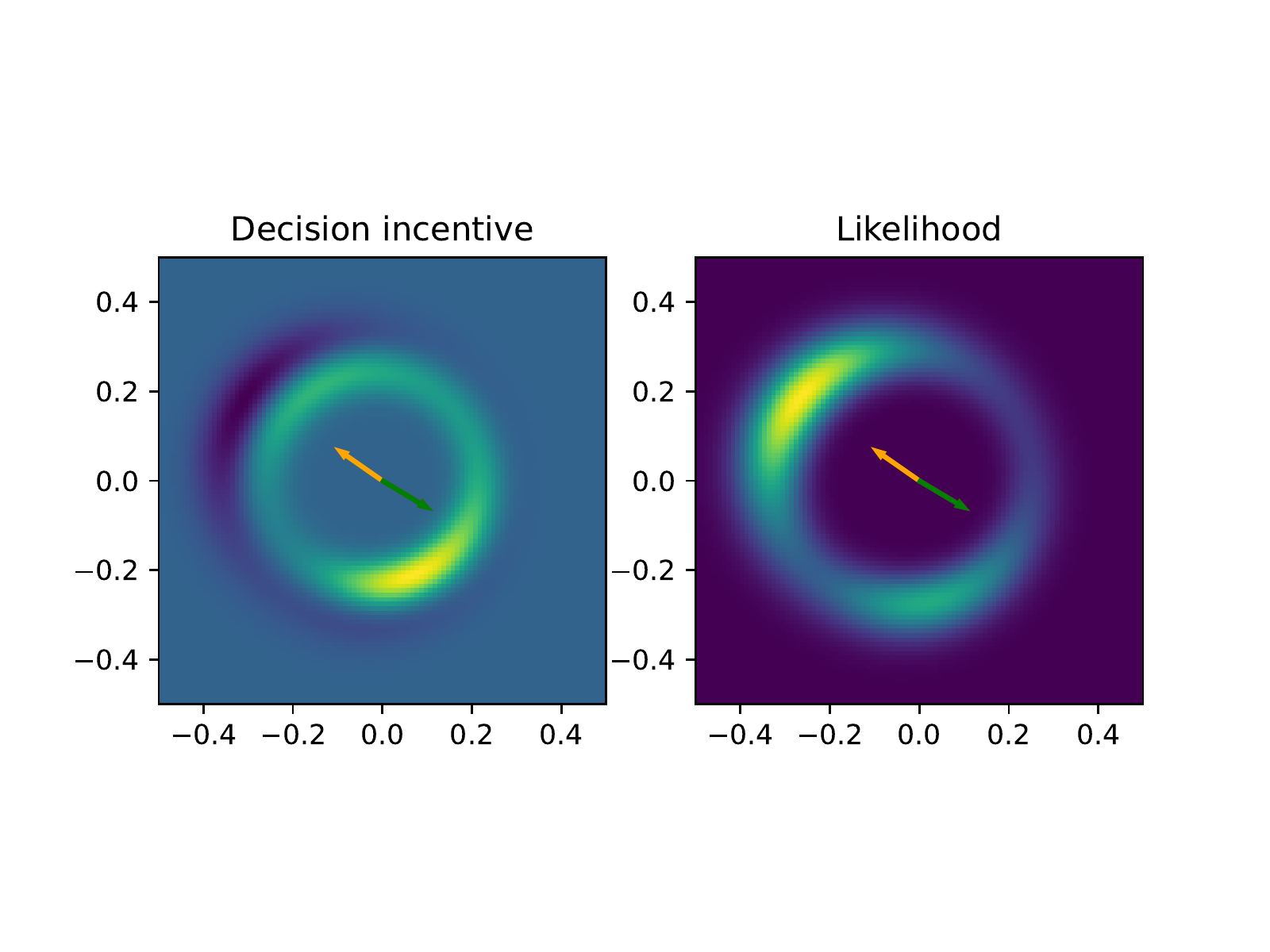}
     \includegraphics[scale=0.48, trim = {0.6cm 50 1 50}, clip]{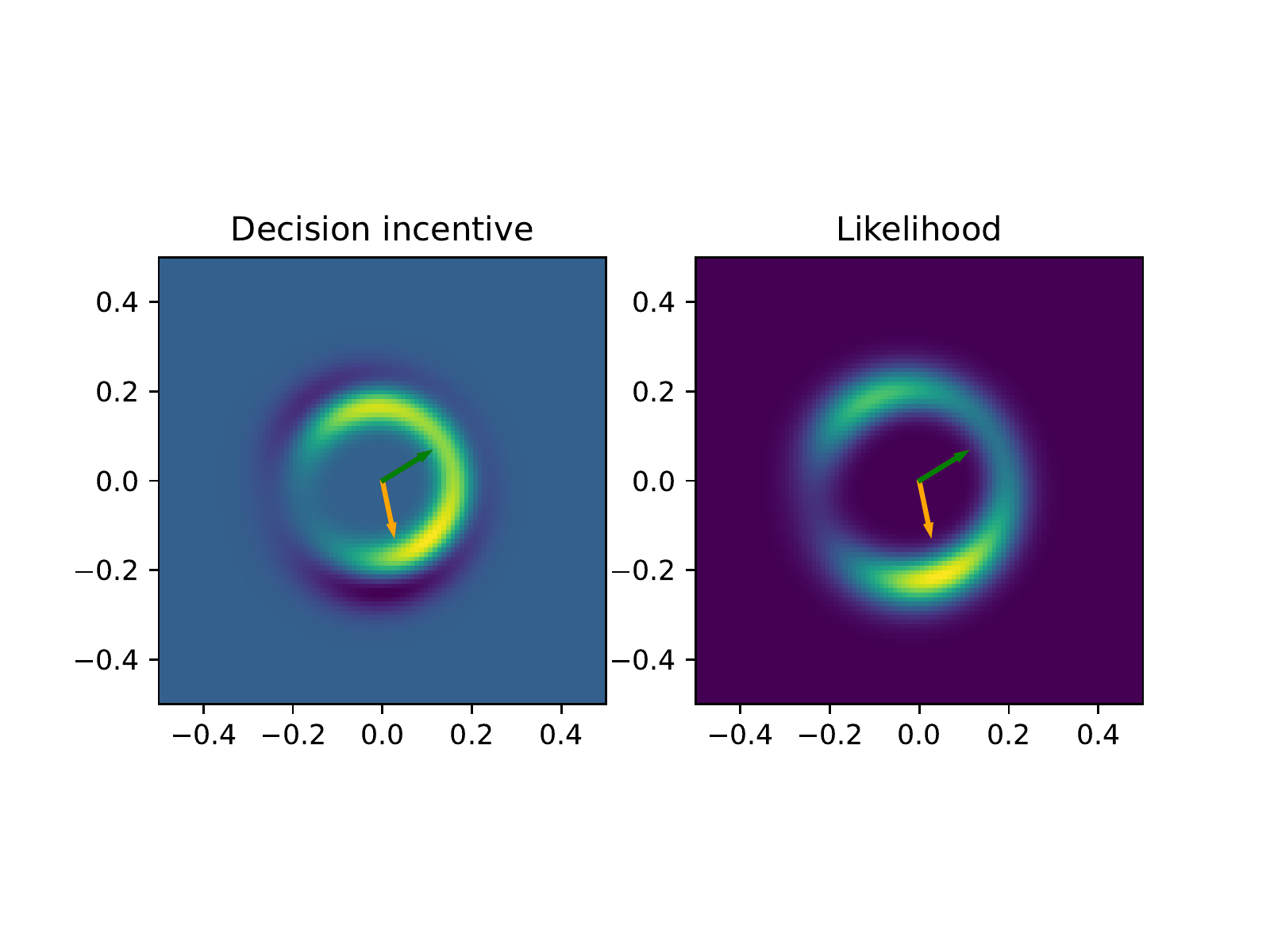}
\caption{Some more examples of likelihood configurations $p$, and corresponding decision incentives $q$, see Eq. \eqref{infotaxis a} and following discussion. The orange arrow corresponds to the maximum likelihood direction, while the green arrow to the infotaxis direction. The particular configurations were selected from a long steady-state realization with parameters $a=0.01$, $\lambda=2$, $v=0.01$, $D=2.5 \cdot 10^{-4}$, $R_\mathrm{min}=0.03$, $R_\mathrm{max}=0.87$.}
	\label{shapes}
\end{figure}
\end{center}

\end{widetext}

\bibliographystyle{unsrt}
\bibliography{Infotaxis_notes.bib}

\end{document}